\documentclass{jfm}
\usepackage[T1]{fontenc}
\usepackage[latin1]{inputenc}
\usepackage[UKenglish]{babel}
\usepackage{graphicx}
\usepackage{psfrag}
\usepackage{epstopdf}
\usepackage[labelformat=simple]{subcaption}

\usepackage[dvipsnames]{xcolor}
\usepackage{booktabs}
\usepackage{multirow}
\usepackage{marginnote}

\usepackage{natbib}
\usepackage{amsmath,amsfonts,amssymb,wasysym}

\usepackage{tikz}
\usetikzlibrary{shapes.misc}

\graphicspath{{./}
              {img/}
              {img/spectra_compare_small/}
              {img/drag_breakdown/}
              {img/R_stress_breakdown/}
              {img/snapshots_decomposition/}
              {img/one_d_cut/}}

\ifCUPmtlplainloaded \else
  \checkfont{eurm10}
  \iffontfound
    \IfFileExists{upmath.sty}
      {\typeout{^^JFound AMS Euler Roman fonts on the system,
                   using the `upmath' package.^^J}%
       \usepackage{upmath}}
      {\typeout{^^JFound AMS Euler Roman fonts on the system, but you
                   dont seem to have the}%
       \typeout{`upmath' package installed. JFM.cls can take advantage
                 of these fonts,^^Jif you use `upmath' package.^^J}%
      }
  \else
  \fi
\fi

\ifCUPmtlplainloaded \else
  \checkfont{msam10}
  \iffontfound
    \IfFileExists{amssymb.sty}
      {\typeout{^^JFound AMS Symbol fonts on the system, using the
                `amssymb' package.^^J}%
       \usepackage{amssymb}%
         
         \let\geq=\geqslant
      }{}
  \fi
\fi

\ifCUPmtlplainloaded \else
  \IfFileExists{amsbsy.sty}
    {\typeout{^^JFound the `amsbsy' package on the system, using it.^^J}%
     \usepackage{amsbsy}}
    {\providecommand\boldsymbol[1]{\mbox{\boldmath $##1$}}}
\fi

\newcommand{\mylab}[3]{\raisebox{#2}[0mm][0mm]{\makebox[0mm][l]{\hspace*{#1}#3}}}

\newcommand\yp {\ensuremath{y^+}}
\renewcommand\k  {\ensuremath{k}}
\newcommand\kp {\ensuremath{k^+}}
\newcommand\ks {\ensuremath{k_s}}
\newcommand\ksp{\ensuremath{k^+_{s}}}
\newcommand\ksi{\ensuremath{k_{s_\infty}}}
\newcommand\ksip{\ensuremath{k^+_{s_\infty}}}
\newcommand\DCf{\ensuremath{\Delta C_f}}

\newcommand\DUp{\ensuremath{\Delta U^+}}
\newcommand\Du{\ensuremath{\ell_{u}}}
\newcommand\Duv{\ensuremath{\ell_{uv}}}
\newcommand\Dj{\ensuremath{\ell_{J}}}
\newcommand\Dup{\ensuremath{\ell_{u}^+}}
\newcommand\Duvp{\ensuremath{\ell_{uv}^+}}
\newcommand\Djp{\ensuremath{\ell_{J}^+}}

\newcommand\Utips {\ensuremath{U_0}}
\newcommand\Utipsp{\ensuremath{U_0^+}}
\newcommand\vtipsp{\ensuremath{v_t^{\prime +}}}
\newcommand\protu {\ensuremath{\ell_U  }}
\newcommand\protup{\ensuremath{\ell_U^+}}

\newcommand\Retau{\ensuremath{\text{Re}_\tau}}
\newcommand\Ret{\ensuremath{\text{Re}_\tau}}
\newcommand\utau{\ensuremath{u_\tau}}

\newcommand\ReStress[1][]{\ensuremath{{-\langle uv \rangle}}}
\newcommand\ReStressp[1][]{\ensuremath{{-\langle uv \rangle}^+_{#1}}}
\newcommand\minusReStressp[1][]{\ensuremath{{\langle uv \rangle}^+_{#1}}}
\newcommand\dsp{\ensuremath{\delta^+_s}}
\newcommand\drp{\ensuremath{\delta^+_r}}

\newcommand\hrp{\ensuremath{h^+_r}}
\newcommand\Tterm{\ensuremath{\mathcal{T}}}

\newcommand\Lx {\ensuremath{L_x}}
\newcommand\Lz {\ensuremath{L_z}}
\newcommand\Lxp{\ensuremath{L_x^+}}
\newcommand\Lzp{\ensuremath{L_z^+}}
\newcommand\sx {\ensuremath{s_x}}
\newcommand\sz {\ensuremath{s_z}}

\newcommand\uT{\ensuremath{u_{BT}}}
\newcommand\vT{\ensuremath{v_{BT}}}
\newcommand\wT{\ensuremath{w_{BT}}}

\newcommand\meanU{\ensuremath{U}}

\newcommand\tilV{\ensuremath{\widetilde{v}}}
\newcommand\tilW{\ensuremath{\widetilde{w}}}

\newcommand\uRu{\ensuremath{u_{RC,u}}}
\newcommand\uRw{\ensuremath{u_{RC,w}}}
\newcommand\uRv{\ensuremath{u_{RC,v}}}
\newcommand\vRu{\ensuremath{v_{RC,u}}}
\newcommand\vRw{\ensuremath{v_{RC,w}}}
\newcommand\vRv{\ensuremath{v_{RC,v}}}
\newcommand\wRu{\ensuremath{w_{RC,u}}}
\newcommand\wRw{\ensuremath{w_{RC,w}}}
\newcommand\wRv{\ensuremath{w_{RC,v}}}

\newcommand\aaa{\textit{a}}
\newcommand\bbb{\textit{b}}
\newcommand\ccc{\textit{c}}
\newcommand\ddd{\textit{d}}

\newcommand\drawline[2]{\raise 2.5pt\vbox{\hrule width #1pt height #2pt}}
\newcommand\spacce[1]{\hskip #1pt}
\newcommand\bdash  {\hbox{\drawline{4}{.8}}}
\newcommand\dashed {\bdash\spacce{2}\bdash\nobreak\ }
\newcommand\bdot   {\hbox{\drawline{1}{.8}\spacce{2}}}
\newcommand\dotted {\hbox{\leaders\bdot\hskip 24pt}\nobreak\ }

\newcommand\chndot {\hbox {\drawline{4}{.8}\spacce{2}\drawline{1}{.8}\spacce{2}\drawline{4}{.8}}\nobreak\ }

\definecolor{cblue}{rgb}{0,0,1}
\definecolor{clightblue}{rgb}{.4,.4,1}
\definecolor{cred}{rgb}{1,0,0}
\definecolor{clightred}{rgb}{1,.4,.4}
\definecolor{cgreen}{rgb}{.3,.7,.2}
\definecolor{clightgreen}{rgb}{.58,.82,.58}
\definecolor{cpurple}{rgb}{.5,0,.5}
\definecolor{clightpurple}{rgb}{.7,.4,.7}
\definecolor{cblack}{rgb}{0,0,0}
\definecolor{clightblack}{rgb}{.4,.4,.4}

\newcommand\linefactory[1]{\begin{tikzpicture}\draw [line width=1.0pt,#1] (0,0) -- (14pt,0);\end{tikzpicture}}
\newcommand\blueline  {\raisebox{.2em}{\linefactory{cblue}}}
\newcommand\redline   {\raisebox{.2em}{\linefactory{cred}}}
\newcommand\purpleline{\raisebox{.2em}{\linefactory{cpurple}}}
\newcommand\greenline {\raisebox{.2em}{\linefactory{cgreen}}}
\newcommand\blackline {\raisebox{.2em}{\linefactory{black}}}

\newcommand\shortredline{\begin{tikzpicture}\draw [line width=1.0pt,cred] (0,0) -- (6pt,0);\end{tikzpicture}}
\newcommand\shortblueline{\begin{tikzpicture}\draw [line width=1.0pt,cblue] (0,0) -- (6pt,0);\end{tikzpicture}}
\newcommand\shortblackline{\begin{tikzpicture}\draw [line width=1.0pt,black] (0,0) -- (6pt,0);\end{tikzpicture}}
\newcommand\redlinedash{\raisebox{.2em}{\shortredline\hspace{2pt}\shortredline}}
\newcommand\bluelinedash{\raisebox{.2em}{\shortblueline\hspace{2pt}\shortblueline}}
\newcommand\blacklinedash{\raisebox{.2em}{\shortblackline\hspace{2pt}\shortblackline}}

\newcommand\circlefactory[2]{\begin{tikzpicture}\filldraw[fill=#1,draw=#2,line width=1pt] circle (2.5pt);\end{tikzpicture}\,}
\newcommand\bluecircle  {\circlefactory{clightblue}{cblue}}

\newcommand\blackcircle {\circlefactory{clightblack}{cblack}}

\newcommand\squarefactory[2]{\begin{tikzpicture}\filldraw[fill=#1,draw=#2,line width=1pt] (0,0) rectangle (5pt,5pt);\end{tikzpicture}\,}

\newcommand\redsquare   {\squarefactory{clightred}{cred}}

\newcommand\blacksquar  {\squarefactory{clightblack}{cblack}}

\newcommand\trianglefactory[2]{\begin{tikzpicture}\filldraw[fill=#1,draw=#2,line width=1pt] (5pt,0) -- (0,0) -- (2.5pt,5pt) -- (5pt,0) -- (0,0);\end{tikzpicture}\,}

\newcommand\purpletriangle{\trianglefactory{clightpurple}{cpurple}}

\newcommand\blacktri      {\trianglefactory{clightblack}{cblack}}
\newcommand\emptyblacktri {\trianglefactory{white}{cblack}}

\newcommand\diamondfactory[2]{\begin{tikzpicture}\filldraw[fill=#1,draw=#2,line width=1pt] (0,0) -- (-2.4pt,3pt) -- (0,6pt) -- (2.4pt,3pt) -- (0,0) -- (-2.4pt,3pt);\end{tikzpicture}\,}

\newcommand\greendiamond {\diamondfactory{clightgreen}{cgreen}}
\newcommand\blackdiamond {\diamondfactory{clightblack}{cblack}}

\newcommand\emptycirclefactory[1]{\begin{tikzpicture}\filldraw[fill=white,draw=#1,line width=1pt] circle (2.5pt);\end{tikzpicture}\,}
\newcommand\emptyblackcircle {\emptycirclefactory{black}}

\tikzset{cross/.style={cross out, draw=black, minimum size=2*(#1-\pgflinewidth), inner sep=0pt, outer sep=0pt,thick},
cross/.default={1pt}}
\newcommand\blackcross  {\begin{tikzpicture}\draw (0.25,.25) node[cross=2.8pt,rotate=0,black]{};\end{tikzpicture}\,}

\newcommand\blackplus  {\begin{tikzpicture}\draw (0.25,.25) node[cross=2.8pt,rotate=45,black]{};\end{tikzpicture}\,}

\title[Modulation in the transitionally rough regime]{Modulation of near-wall turbulence in the transitionally rough regime}

\author[N. Abderrahaman-Elena, C. T. Fairhall and R. Garc\'{i}a-Mayoral]%
{Nabil Abderrahaman-Elena$^1$, Chris T. Fairhall$^1$ and Ricardo Garc\'{i}a-Mayoral$^1$
\thanks{Email address for correspondence: r.gmayoral@eng.cam.ac.uk}\ns}

\affiliation{$^1$Engineering Department, University of
Cambridge, Trumpington Road, Cambridge CB2 1PZ, UK}

\pubyear{?}
\volume{?}
\pagerange{?--?}
\date{?; revised ?; accepted ?. - To be entered by editorial office}
\begin{document}

\maketitle

\begin{abstract}
Direct numerical simulations of turbulent channels with rough walls are conducted in the transitionally rough regime.
The effect that roughness produces on the overlying turbulence is studied using a modified triple decomposition of the flow.
This decomposition separates the roughness-induced contribution from the background turbulence, with the latter essentially free of any texture footprint.
For small roughness, the background turbulence is not significantly altered, but merely displaced closer to the roughness crests,
with the change in drag being proportional to this displacement.
As the roughness size increases, the background turbulence begins to be modified, notably by the increase of energy for short, wide wavelengths, which is consistent with the appearance of a shear-flow instability of the mean flow.
A laminar model is presented to estimate the roughness-coherent contribution, as well as the displacement height and the velocity at the roughness crests.
Based on the effects observed in the background turbulence, the roughness function is decomposed into different terms to analyse different contributions to the change in drag, laying the foundations for a predictive model.
\end{abstract}

\section{Transitionally rough regime}\label{sec:introduction}

\begin{figure}
  \centering
  \includegraphics[width=70mm]{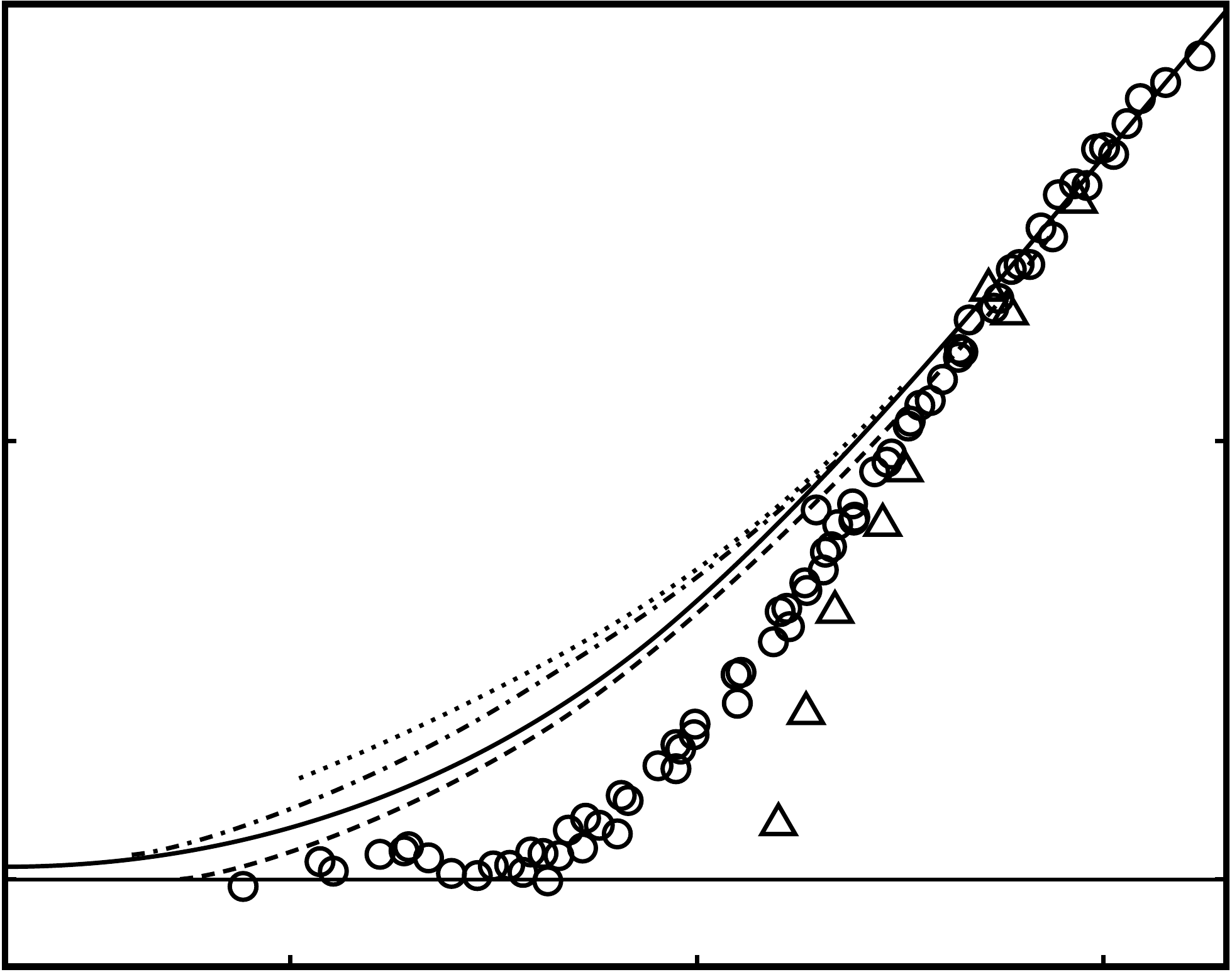}
  \mylab{-80.0mm}{43mm}{$\DUp$}%
  \mylab{-75.0mm}{53.4mm}{$10$}%
  \mylab{-74.0mm}{29.0mm}{$5$}%
  \mylab{-74.0mm}{ 4.5mm}{$0$}%
  \mylab{-42mm}{-8mm}{$\ksip$}%
  \mylab{-57.0mm}{-4mm}{$10^0$}%
  \mylab{-34.0mm}{-4mm}{$10^1$}%
  \mylab{-11.0mm}{-4mm}{$10^2$}%
  \vspace*{2.2em}
  \caption{Roughness function \DUp\ in the transitionally rough regime as a function of \ksip. \protect\emptyblackcircle, uniform sand \citep{Nikuradse1933}; \protect\emptyblacktri, uniform packed spheres \citep{Ligrani1986}; \dotted, galvanized iron; \dashed, tar-coated cast iron; \chndot, wrought-iron; \protect\blackline, interpolation \citep{Colebrook1939}. Adapted from \citet{Jim2004}.}
  \label{fig:transitional_roughness_adapted}
\end{figure}

Although one of the oldest problems in fluid mechanics, roughness in turbulent flows remains an active area of research.
The seminal works by \citet{Nikuradse1933} and \citet{Colebrook1937}, offering a systematic study of rough surfaces, set the foundations for the field.
Based upon Nikuradse's results on pipes roughened with sand of equal grain size, \citet{Schlichting1936} introduced the concept of equivalent sand grain size, \ks, in order to classify rough surfaces.
Surfaces producing the same friction as a particular sand grain size would share the same \ks.
However, there are two important shortcomings in this strategy.
Firstly, as stated by \citet{Bradshaw2000}, it ``simply defines a useful common currency for roughness size---like paper money, valueless in itself but normally acceptable as a medium of exchange'', i.e. \ks\ is merely a classifier, an a posteriori parameter that is of limited use for prediction purposes.
Furthermore, \citet{Colebrook1939} observed that, despite sharing the same value of \ks\ in the fully rough regime, different surfaces vary in how they depart from the hydraulically smooth regime.
In figure~\ref{fig:transitional_roughness_adapted}, these differences are highlighted by using \ksip \citep{Jim2004}.
In this figure, a roughness lengthscale \ksi\ is chosen so that in the fully rough regime the roughness function \DUp\ depends only on \ksip.
The figure highlights that, while such lengthscale can collapse the results in the fully rough regime, differences can still be observed in the transitionally rough regime.
It is also worth nothing that \ksi\ can only be obtained a posteriori from experimental or numerical results with varying \ksip.
Many studies in the recent years have aimed to find a suitable combination of parameters to describe roughness surfaces and predict their friction.
Some authors have explored the effective slope, the solidity, as well as different moments of the roughness height, mainly the mean, the standard deviation and the skewness \citep{Flack2010}.
A universal governing parameter, however, has not been found, and the changes in the flow that lead to the increase in friction have not been fully understood.
The approach taken in this work is to understand the modifications produced on the flow by the roughness texture.
As stated by \citet{Marusic2010a}, ``without further theoretical advances, there is a risk of needing a catalogue of roughness results''.

In the hydraulically smooth regime, roughness is of small size and has a negligible effect on the flow, causing no change in terms of wall friction. 
As a result, the flow is effectively equivalent to canonical smooth-wall turbulence.
In the fully rough regime, on the other hand, roughness dominates.
The friction coefficient becomes independent of viscosity and the roughness size determines the skin friction drag.
Between these two regimes, in the so-called transitionally rough regime, both the viscous and the rough effects have comparable importance.
\citet{Nikuradse1933} observed that sand grain roughness was hydraulically smooth for $\ksp < 5$, fully rough for $\ksp > 70$, and transitional between those extremes.
However, \citet{Flack2010} compile experimental results for typologies of roughness other than sand grain roughness, finding that the limits vary considerably.
In particular, the literature shows the transitionally rough regime spanning $1.4$--$15<\ksp<18$--$70$.
The lower bound of this regime has classically been seen as a threshold below which the surface completely behaves as hydraulically smooth.
However, \citet{Bradshaw2000} proposes a gradual transition between these regimes, without a defined, hard boundary.
A recent work by \citet{Thakkar2018} also seems to point in this direction.
The interest in transitional roughness resides in understanding the effects that small surface texture starts having on the flow; effects that, as size increases, will lead to the departure from its smooth-wall behaviour and eventually to the fully rough regime.
This interest is also growing amongst the high-Reynolds-number community \citep{Marusic2010a}.
Apparently, smooth surfaces, with a well-controlled micro-texture, may eventually enter the transitionally rough regime as viscous scales become sufficiently small for increasing Reynolds number.

Far enough over conventionally rough walls, the mean velocity profile exhibits the same logarithmic region found over smooth walls.
Effectively, roughness only modifies the intercept of the logarithmic velocity profile, while the K\'{a}rm\'{a}n constant, $\kappa$, and the wake function are unaffected \citep{Nikuradse1933,Clauser1956}.
In the logarithmic layer, the mean velocity profile, $U$, can then be expressed as
\begin{equation}
  U^+ = \kappa^{-1}\ln \left( y^+ \right) + B + \DUp = U^+_0 + \DUp,
  \label{eq:log_profile}
\end{equation}
where $B$ is the smooth-wall intercept of the logarithmic velocity profile.
The roughness function, \DUp\ \citep{Hama1954}, depends on the roughness texture and its size.
The subscript~${}_0$ denotes smooth-wall flow quantities.
The superscript~${}^+$ indicates scaling in wall-units, with the friction velocity, \utau, and the kinematic viscosity, $\nu$.
The change in friction coefficient, \DCf, can be directly related to \DUp. At the centreline of a channel or boundary layer thickness, denoted by the subscript $\delta$, equation~\eqref{eq:log_profile} becomes
\begin{align}
  \left(2/C_f\right)^{1/2} = U^+_{\delta} = U^+_{\delta_0} + \DUp.
\end{align}

\begin{figure}
\centering
\vspace*{0.8cm}
\includegraphics[width=1.0\textwidth]{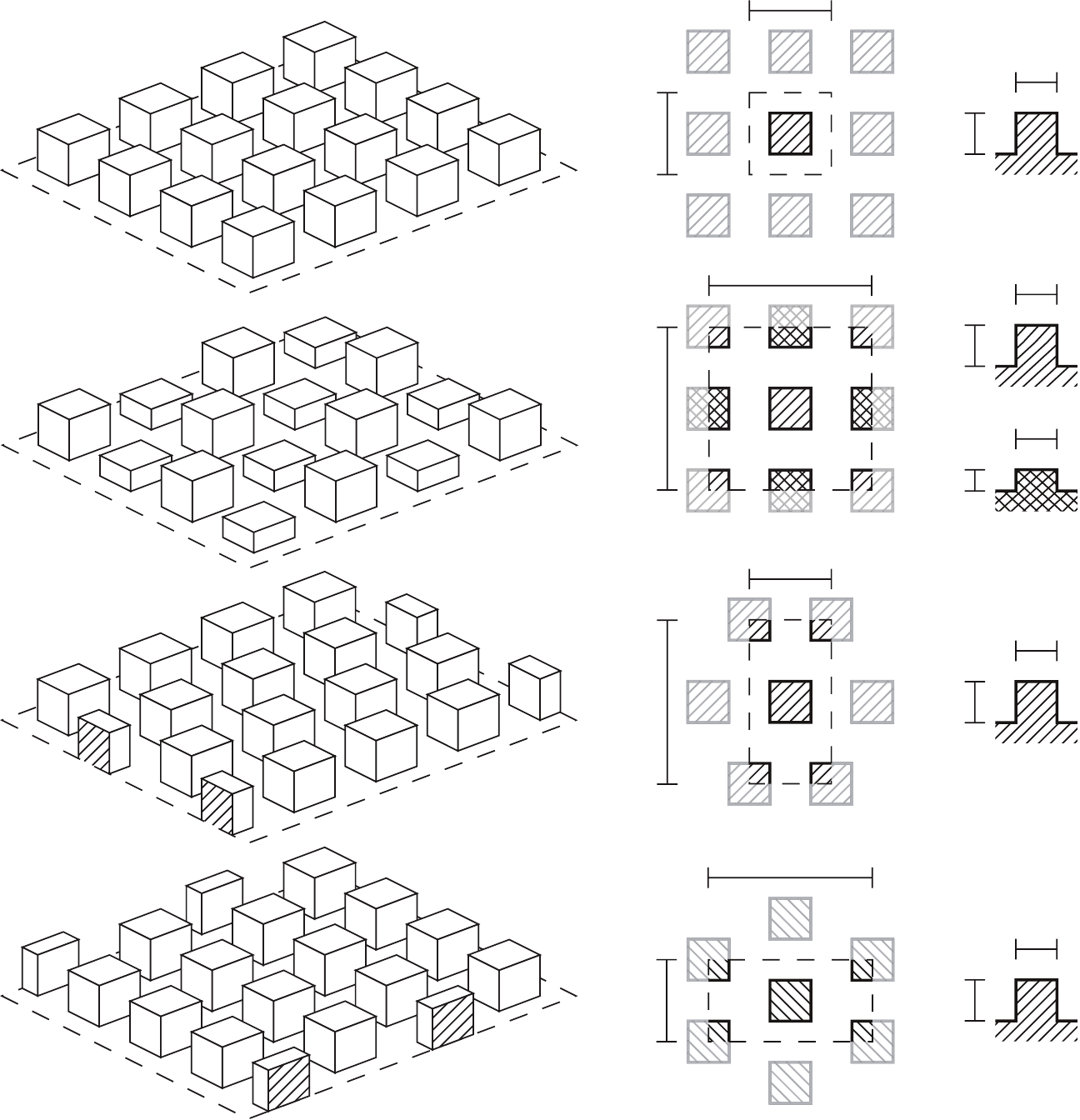}%
\mylab{-1.000\textwidth}{13.4cm}{\aaa)}%
\mylab{-1.000\textwidth}{ 9.8cm}{\bbb)}%
\mylab{-1.000\textwidth}{ 6.2cm}{\ccc)}%
\mylab{-1.000\textwidth}{ 2.8cm}{\ddd)}%
\mylab{-0.278\textwidth}{13.95cm}{\sx}
\mylab{-0.278\textwidth}{10.52cm}{\sx}
\mylab{-0.278\textwidth}{6.85cm}{\sx}
\mylab{-0.278\textwidth}{3.15cm}{\sx}
\mylab{-0.420\textwidth}{12.25cm}{\sz}
\mylab{-0.420\textwidth}{8.85cm}{\sz}
\mylab{-0.420\textwidth}{5.15cm}{\sz}
\mylab{-0.420\textwidth}{1.45cm}{\sz}
\mylab{-0.125\textwidth}{12.20cm}{\k}
\mylab{-0.125\textwidth}{9.58cm}{\k}
\mylab{-0.141\textwidth}{7.88cm}{$\tfrac{1}{2}$\k}
\mylab{-0.125\textwidth}{5.10cm}{\k}
\mylab{-0.125\textwidth}{1.38cm}{\k}
\mylab{-0.046\textwidth}{13.08cm}{\k}
\mylab{-0.046\textwidth}{10.46cm}{\k}
\mylab{-0.046\textwidth}{8.64cm}{\k}
\mylab{-0.046\textwidth}{6.00cm}{\k}
\mylab{-0.046\textwidth}{2.25cm}{\k}
\caption{Schematics of the roughness patterns studied. (\aaa) Collocated; (\bbb) Collocated with two heights; (\ccc) Spanwise-staggered and (\ddd) Streamwise-staggered. The mean flow is from bottom left to top right. Left panels, three-dimensional representation of the roughness surface; central panels, top view of a unit element of the pattern; right panels, side view of an isolated post.}%
\label{fig:sketch_roughness_geometries}
\end{figure}
In this work we explore the transitionally rough regime. 
The ultimate goal is to understand and predict the effect of transitionally rough surfaces on the flow.
The range of roughness sizes covered is therefore small, but of more general relevance as this regime marks the onset of the rough behaviour.
To explore it, we conduct a campaign of direct numerical simulations (DNSs) of turbulent channels.
The channels are symmetric, with roughness on both walls.
Four different geometries are considered, as depicted in figure~\ref{fig:sketch_roughness_geometries}.
To explore the behaviour of each geometry, direct numerical simulations are conducted for different values of \k, while their shape is kept fixed, also resulting in constant solidity and effective slope.
These surfaces exhibit the classic k-roughness behaviour, which is characteristic of most three-dimensional rough surfaces \citep{Jim2004}.

This paper is organised as follows.
In \S\ref{sec:methodology} the details of the numerical method are presented.
A novel decomposition of the flow is presented in \S\ref{sec:flow_decomp} and it is used in \S\ref{sec:rough_comp} and \S\ref{sec:the_turbulent_component} to study the two components of the velocity.
In \S\ref{sec:kelvin_helmholtz} the appearance of a shear-flow instability is discussed, and the relationship between the roughness function and the virtual origin of turbulence is analysed in \S\ref{sec:stress_breakdown}.
The key conclusions are discussed in \S\ref{sec:conclusions}.

\section{Methodology}\label{sec:methodology}

The numerical experiments are conducted in an incompressible, turbulent channel with roughness on the top and bottom walls.
The domain is periodic in the wall-parallel directions. 
The channel half-height is $\delta$ measured from the roughness crests, and the length and width are $\Lx = 2\pi\delta$ and $\Lz = \pi\delta$, respectively.
The streamwise, wall-normal and spanwise coordinates are $x$, $y$ and $z$, with $u$, $v$ and $w$ the corresponding components of the velocity $\boldsymbol{u}$.

The numerical method is adapted from that of \citet{GarM2011} for riblet channels to simulate fully three-dimensional roughness and is briefly summarised here.
The temporal integrator is a fractional-step method combined with a three--sub-step Runge-Kutta and with pressure correction at the final sub-step only \citep{Le1991,Perot1993}.
The spatial discretisation is pseudo-spectral.
The two periodic directions, $x$ and $z$, are discretised using Fourier series, while the wall-normal direction, $y$, is discretised using a second order finite difference scheme in a collocated grid.
The chequerboard effect, typical of collocated grids \citep{Ferziger2002}, is addressed using a quasi-divergence-free formulation \citep{Nordstrom2007,GarM2011}.
Near the walls, the high velocity gradients and the roughness texture require a high spatial resolution for the flow to be correctly resolved, while for the core of the channel, a coarser, less costly resolution would suffice.
To alleviate the computational cost of the method while still resolving all scales, two strategies are implemented.
Along the wall-normal direction the finite-difference discretisation allows for the grid to be stretched, with a higher density of points near the walls.
Above the roughness, the wall-normal grid spacing is $\Delta y^+_\text{min} \approx 0.3$ at the roughness crests, and progressively grows to $\Delta y^+_\text{max} \approx 3.1$ at the centre of the channel.
Within the roughness region, i.e.\ between roughness crests and troughs, the grid spacing is $\Delta y^+ \approx 0.3$.
Additionally, to increase the $x$- and $z$-resolutions near the walls, the domain is vertically divided into three blocks.
This multi-block technique allows a high number of grid points to be set near the walls, with a coarser resolution in the block resolving the core of the channel.
In this latter region the resolution is set to resolve all turbulent scales.
In the blocks containing the walls, it is also necessary to resolve the flow around the roughness elements, which requires a higher resolution for small \kp.
The resolution in the central block is $\Delta_{x_c}^+ \approx 6$ and $\Delta_{z_c}^+ \approx 3$ along the $x$ and $z$ directions respectively, while in the two wall blocks a finer resolution is used to represent each roughness element, with at least 12 points along $x$ and $z$. 
The roughness elements are resolved using a direct-forcing immersed boundary method \citep{Mohd-Yusof1997,Fadlun2000,Iaccarino2003}.
The unit element is a cube of side \k, repeated along $x$ and $z$ on both walls.
These unit elements are described using $12\times12$ points for the cases with $\kp \lesssim 9$, $24\times24$ points for the cases with $12 \lesssim \kp \lesssim 25$, and $48\times48$ points for the case with $\kp \approx 36$.
This choice results from the need to solve both the turbulent scales as well as the flow around the roughness elements while keeping a moderate computational cost.
Note that, although the resolution for the smallest \kp\ values is marginal, the simulations carried out by \citet{Thakkar2018} on random roughness, performed using a 12 point-per-element discretisation, are in good agreement with experimental results.

The simulations are conducted at constant mass flow rate, which is adjusted to achieve a friction Reynolds number $\Ret \approx 185$ for all simulations.
\Ret\ is computed using $\delta' = \delta + \protu$, where \protu\ is the virtual origin of the mean velocity profile, i.e. where the mean velocity profile tends to zero when extrapolated from $y > 0$.
The friction velocity $u_\tau$ is obtained by extrapolating the total shear stress to $\delta'$.
Statistically converged initial conditions are obtained by simulating and discarding the flow during $10\, \delta/\utau$.
Statistics are then collected over at least $15\, \delta/\utau$. 
Relevant parameters and results of the simulations are given in table~\ref{tab:simulations_data}.
\begin{table}
\centering
\def~{\hphantom{0}}
\begin{tabular}{@{\hspace{1em}}c@{\hspace{1em}}lccrrrrrrrrrr}
  & Case & $\sx/\k$ & $\sz/\k$ & $\delta/\k$ & \kp    & \ksp   & \DUp  & \Retau  & \Utipsp & \protup & \Dup  & \Djp   & \Duvp \\ \midrule
Smooth Channel
  & SC   & $-$      & $-$      & $ - $       & $ 0.0$ & $ 0.0$ & $0.0$ & $183.9$ & $0.0$  & $0.0$   & $0.0$ & $0.0$  & $0.0$ \\ \midrule
\multirow{7}{*}{Collocated}
  & C06  & $2$      & $2$      & $30.6$      & $ 6.0$ & $ 9.6$ & $0.5$ & $183.9$ & $0.5$  & $0.5$   & $0.5$ & $0.6$  & $1.2$ \\
  & C09  & $2$      & $2$      & $20.4$      & $ 8.8$ & $14.1$ & $0.7$ & $180.8$ & $0.7$  & $0.7$   & $0.7$ & $0.7$  & $1.5$ \\
  & C12  & $2$      & $2$      & $15.4$      & $11.7$ & $18.7$ & $1.5$ & $180.0$ & $1.1$  & $1.2$   & $1.2$ & $1.2$  & $3.2$ \\
  & C15  & $2$      & $2$      & $12.3$      & $14.4$ & $23.1$ & $2.4$ & $178.7$ & $1.3$  & $1.5$   & $1.6$ & $1.6$  & $4.5$ \\
  & C18  & $2$      & $2$      & $10.3$      & $17.4$ & $27.8$ & $3.5$ & $179.0$ & $1.5$  & $1.9$   & $1.9$ & $2.3$  & $6.3$ \\
  & C24  & $2$      & $2$      & $ 7.7$      & $22.5$ & $36.0$ & $4.7$ & $174.5$ & $1.6$  & $2.5$   & $2.6$ & $3.8$  & $8.4$ \\
  & C36  & $2$      & $2$      & $ 5.2$      & $35.7$ & $57.2$ & $6.7$ & $186.7$ & $2.1$  & $4.4$   & $4.8$ & $8.8$  & $-$   \\ \midrule 
\multirow{5}{*}{\begin{tabular}[c]{@{}c@{}}Collocated \\two heights\end{tabular}}    
  & CC06 & $4$      & $4$      & $30.7$      & $ 5.8$ & $-$    & $0.8$ & $178.7$ & $1.2$  & $1.2$   & $1.4$ & $1.3$  & $2.1$ \\
  & CC09 & $4$      & $4$      & $20.6$      & $ 8.7$ & $-$    & $1.8$ & $179.8$ & $1.6$  & $1.7$   & $2.1$ & $1.7$  & $3.7$ \\
  & CC12 & $4$      & $4$      & $15.5$      & $11.7$ & $-$    & $3.3$ & $182.6$ & $2.1$  & $2.7$   & $3.0$ & $2.6$  & $6.2$ \\
  & CC15 & $4$      & $4$      & $11.7$      & $15.4$ & $-$    & $4.5$ & $180.1$ & $2.4$  & $3.6$   & $3.4$ & $3.5$  & $8.3$ \\
  & CC18 & $4$      & $4$      & $10.4$      & $17.0$ & $-$    & $4.9$ & $176.7$ & $2.4$  & $3.9$   & $3.5$ & $4.0$  & $9.1$ \\ \midrule
\multirow{4}{*}{\begin{tabular}[c]{@{}c@{}}Spanwise \\staggered\end{tabular}}    
  & SZ06 & $2$      & $4$      & $30.6$      & $ 5.9$ & $-$    & $0.4$ & $183.0$ & $0.5$  & $0.5$   & $0.6$ & $0.6$  & $1.0$ \\
  & SZ09 & $2$      & $4$      & $20.4$      & $ 8.9$ & $-$    & $0.9$ & $182.7$ & $0.7$  & $0.7$   & $0.8$ & $0.8$  & $1.7$ \\
  & SZ12 & $2$      & $4$      & $15.4$      & $11.8$ & $-$    & $1.6$ & $181.8$ & $1.2$  & $1.3$   & $1.3$ & $1.3$  & $3.4$ \\
  & SZ15 & $2$      & $4$      & $11.6$      & $16.2$ & $-$    & $3.2$ & $188.3$ & $1.4$  & $1.8$   & $1.8$ & $2.1$  & $6.0$ \\ \midrule
\multirow{5}{*}{\begin{tabular}[c]{@{}c@{}}Streamwise \\staggered\end{tabular}}    
  & SX06 & $4$      & $2$      & $30.6$      & $ 5.9$ & $-$    & $0.6$ & $181.6$ & $0.5$  & $0.5$   & $0.6$ & $0.5$  & $1.1$ \\
  & SX09 & $4$      & $2$      & $20.4$      & $ 8.8$ & $-$    & $1.2$ & $179.5$ & $0.6$  & $0.6$   & $0.9$ & $0.7$  & $2.0$ \\
  & SX12 & $4$      & $2$      & $15.4$      & $11.9$ & $-$    & $2.6$ & $183.1$ & $1.0$  & $1.1$   & $1.4$ & $1.2$  & $4.2$ \\
  & SX15 & $4$      & $2$      & $11.5$      & $16.0$ & $-$    & $4.1$ & $185.3$ & $1.2$  & $1.4$   & $1.7$ & $1.6$  & $6.0$ \\
  & SX18 & $4$      & $2$      & $10.3$      & $17.6$ & $-$    & $4.3$ & $180.6$ & $1.2$  & $1.5$   & $1.8$ & $1.8$  & $6.2$ \\ \midrule
\end{tabular}
\caption{Main characteristics of the simulations. The roughness-element width and maximum height is $k$, and the streamwise and spanwise pitches of the pattern are \sx\ and \sz. The half-height of the channel, measured from the roughness crests, is denoted by $\delta$. The mean velocity at the roughness crests is \Utips. $\ell^+$ denotes the depth of virtual origins, measured from the tips, as introduced in \S\ref{sec:the_turbulent_component}, where \protup\  refers to the virtual origin of the mean velocity profile; \Dup\ is the virtual origin of the streamwise rms fluctuations; \Djp\ is Jackson's displacement height \citep{Jackson1981}; and \Duvp\ refers to the virtual origin of the Reynolds shear stress.}
\label{tab:simulations_data}
\end{table}

\section{Flow decomposition}\label{sec:flow_decomp}

In the proximity of the wall, roughness induces fluctuations that alter, and even destroy, the near-wall cycle \citep{Jim2004}.
The intensity of these fluctuations decreases exponentially away from the surface, and they are generally considered to be confined within a region near the wall of height $\sim 3k$, known as the roughness-sublayer \citep{Raupach1991,Flack2007}.
In this region the fluctuations induced by roughness cannot be neglected and their effect is noticeable in the flow.
These roughness-coherent fluctuations are of the same order of magnitude as the background turbulence fluctuations, leaving their footprint on the velocity field and the energy spectrum.

\begin{figure}
\centering
\includegraphics[width=1.0\textwidth]{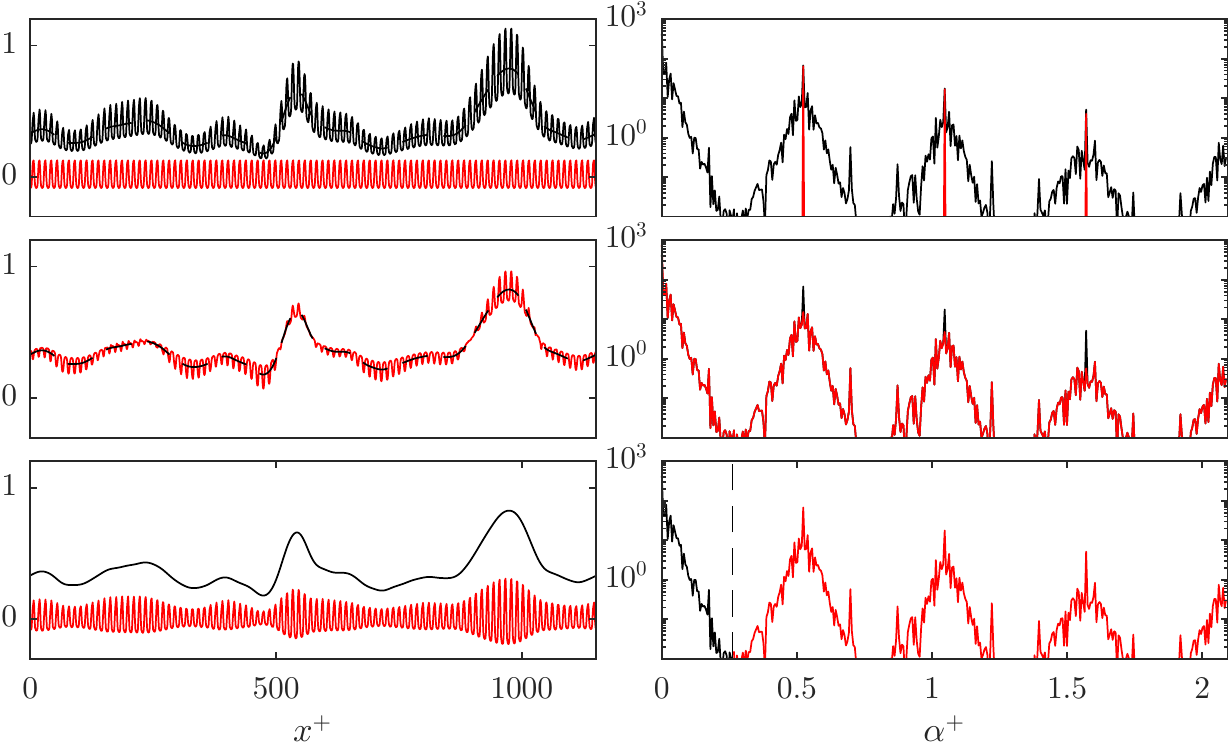}%
\mylab{-0.965\textwidth}{7.60cm}{\aaa.1)}%
\mylab{-0.450\textwidth}{7.60cm}{\aaa.2)}%
\mylab{-0.965\textwidth}{5.18cm}{\bbb.1)}%
\mylab{-0.450\textwidth}{5.18cm}{\bbb.2)}%
\mylab{-0.965\textwidth}{2.75cm}{\ccc.1)}%
\mylab{-0.450\textwidth}{2.75cm}{\ccc.2)}%
\caption{Instantaneous realisation of the streamwise velocity, for case C06, at $y^+ = 0.3$, in a section of constant $z^+$ through the middle of a row of posts.
  (\aaa.1) and (\aaa.2) \protect\blackline, full signal; \protect\blacklinedash, background turbulent component; \protect\redline, ensemble average over time and over the relative position in the periodic unit of texture.
  (\bbb.1) and (\bbb.2) \protect\redline, full signal minus the ensemble average; \protect\blacklinedash, background turbulence.
  (\ccc.1) and (\ccc.2) \protect\blackline, low-pass filtered full signal; \protect\redline, difference of the latter with the full signal.
  The left panels represent the signals in physical space, and the right panels in Fourier space, where $\alpha$ is the wavenumber.
}%
\label{fig:one_d_cut_phys_and_spect_u_h06_bw_JFM}
\end{figure}
\begin{figure}
\centering
\includegraphics[width=0.5\textwidth]{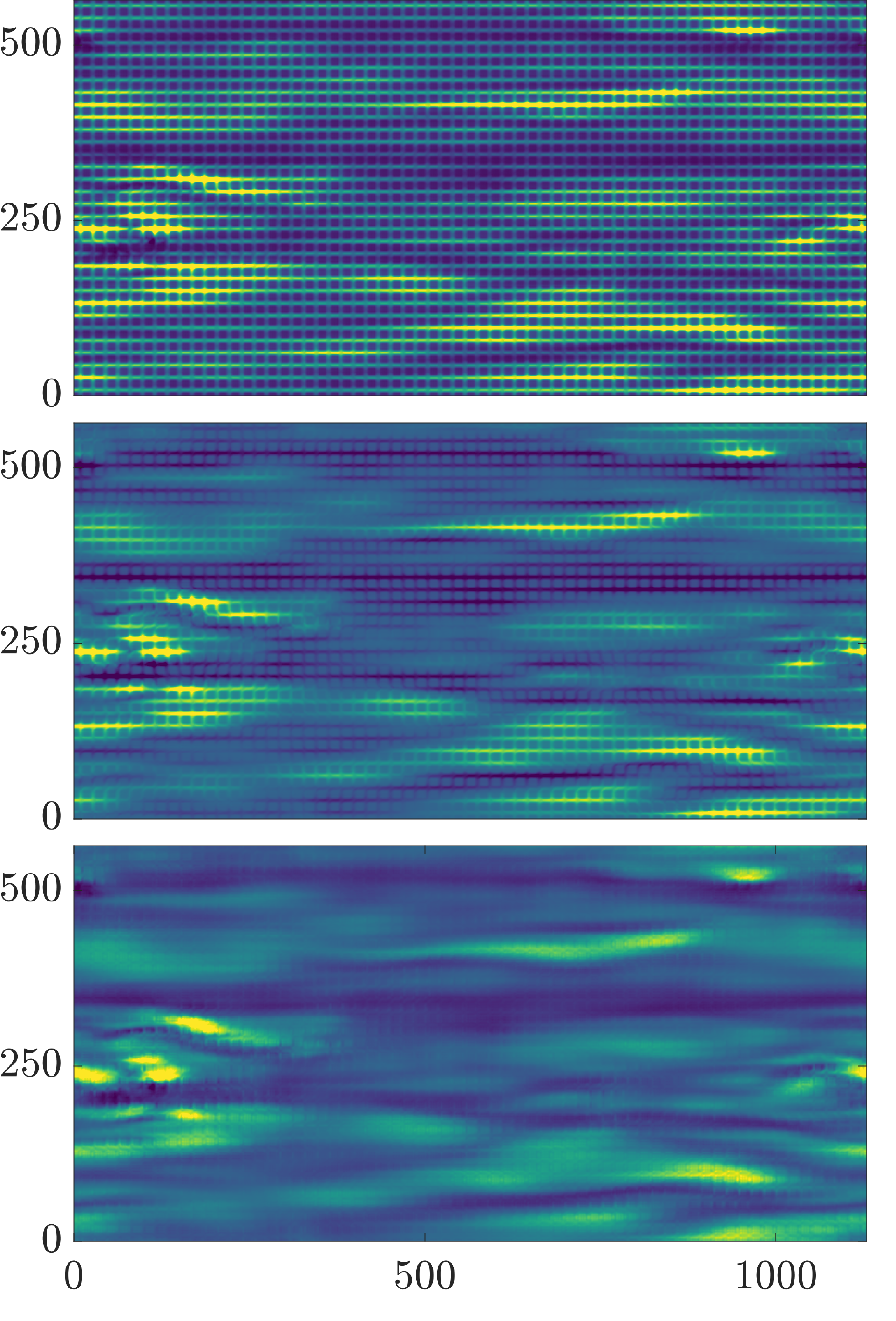}%
\mylab{-0.550\textwidth}{9.80cm}{\aaa)}%
\mylab{-0.550\textwidth}{6.60cm}{\bbb)}%
\mylab{-0.550\textwidth}{3.45cm}{\ccc)}%
\mylab{-0.540\textwidth}{8.50cm}{$L_z^+$}%
\mylab{-0.540\textwidth}{5.30cm}{$L_z^+$}%
\mylab{-0.540\textwidth}{2.15cm}{$L_z^+$}%
\mylab{-0.250\textwidth}{-0.10cm}{$L_x^+$}%
\caption{Instantaneous streamwise velocity at $y^+ \approx 0.3$, for case C09.(\aaa) Full signal. (\bbb) Turbulent contribution obtained using triple decomposition, $u - \meanU - \uRu$. (\ccc) Turbulent contribution obtained using equation~\eqref{eq:decomposition_first_step}.}
\label{fig:snapshots_decomposition_u_h09_bw} 
\end{figure}
In the roughness-sublayer the flow can be thought of as formed of a roughness-coherent component, which is coherent in time and space with the roughness surface, and a background-turbulent component of a chaotic nature.
This is analogous to the decomposition from \citet{Rey1974} into a coherent and a turbulent component.
To illustrate this concept, figure~\ref{fig:one_d_cut_phys_and_spect_u_h06_bw_JFM}(\aaa.1) displays an instantaneous realisation of the streamwise velocity close to the roughness crests. 
We observe two contributions. 
The first, with longer wavelengths, has a characteristic lengthscale of the order of turbulent eddies. 
The second, whose dominant wavelength is that of the roughness texture, smaller than the overlying eddies, can be obtained by ensemble averaging over time and over the relative position in the periodic unit of texture.
If the roughness-coherent component is subtracted from the full velocity, as in figure~\ref{fig:one_d_cut_phys_and_spect_u_h06_bw_JFM}(\bbb.1),  we observe that the footprint of the roughness elements does not completely disappear.
The reason can be explained by means of their corresponding Fourier transform, shown in figures~\ref{fig:one_d_cut_phys_and_spect_u_h06_bw_JFM}(\aaa.2) and~\ref{fig:one_d_cut_phys_and_spect_u_h06_bw_JFM}(\bbb.2).
The ensemble average is composed of the wavelength of the texture and its subharmonics.
The full velocity exhibits these same wavelengths, induced by the texture; however, they are modulated in amplitude and thereby surrounded by energy in the neighbouring wavelengths.
When the ensemble average is subtracted from the full velocity, the most energetic part, which corresponds to the wavelength of the texture and its subharmonics, is removed, but all the energy neighbouring these wavelengths is not, failing to remove the footprint as observed in figure~\ref{fig:one_d_cut_phys_and_spect_u_h06_bw_JFM}(\bbb.1).
In figures~\ref{fig:one_d_cut_phys_and_spect_u_h06_bw_JFM}(\ccc.1) and~\ref{fig:one_d_cut_phys_and_spect_u_h06_bw_JFM}(\ccc.2) we set a threshold to identify what wavelengths correspond to background turbulence and thus what the remainder represents in physical space.
The wavelengths are then divided into long and short ones.
In figure~\ref{fig:one_d_cut_phys_and_spect_u_h06_bw_JFM}(\ccc.1), we observe that the contribution from long wavelengths closely resembles smooth-wall turbulence.
Wavelengths shorter than those of turbulence are similar to the ensemble average, shown in figure~\ref{fig:one_d_cut_phys_and_spect_u_h06_bw_JFM}(\aaa.1), but they appear to be modulated in amplitude by the overlying, long-wavelength signal of the background turbulence.

Following the above analysis, in \citet{Abderrahaman2016} we proposed that the flow over a roughness texture of small size is not simply the sum of the ensemble average plus a turbulent component.
Instead, the roughness-coherent flow is modulated by the overlying turbulence.
For a particular roughness size, inducing a coherent flow \uRu, and with a background turbulence \uT, the instantaneous velocity can intuitively be modelled as 
\begin{equation}
  u \approx \meanU + \uT + \frac{\meanU + \uT}{\meanU} \uRu ,
  \label{eq:decomposition_first_step}
\end{equation}
where \meanU\ is the temporal and $x$-$z$-spatial averaged streamwise velocity.

Equation~\eqref{eq:decomposition_first_step} can also be inferred by considering small roughness in the viscous limit.
In the spirit of \citet{Luc1991}, the vanishingly small roughness limit reduces the problem to a shear driven, purely viscous flow, which results in a self-similar solution that scales with \k\ and with the overlying shear, proportional to $\meanU + \uT$.

The triple decomposition proposed by \citet{Rey1974} is similar to equation~\eqref{eq:decomposition_first_step} except for the amplitude modulation by $\meanU + \uT$.
The triple decomposition has been widely used to characterise the texture-coherent flow in turbulence over a variety of complex surfaces \citep{Choi1993,Jim2001,GarM2011,Jelly2014,Seo2015}.
However, if we are to study the modifications in the background turbulence, the modulation of the roughness-coherent component needs to be accounted for.
Figure~\ref{fig:snapshots_decomposition_u_h09_bw}(\aaa) shows the streamwise velocity within the roughness layer, where the coherent signal of the roughness elements is strong.
If the ensemble average, the roughness-coherent contribution, is subtracted we observe in figure~\ref{fig:snapshots_decomposition_u_h09_bw}(\bbb) that, although reduced, the footprint of the surface texture remains.
However, when we account for the modulation of the coherent flow, using equation \eqref{eq:decomposition_first_step}, the signature of the roughness-coherent flow on the background turbulence is substantially weaker, as shown in figure~\ref{fig:snapshots_decomposition_u_h09_bw}(\ccc) where it is negligible.

\begin{figure}
\centering
\vspace{0.5cm}%
\includegraphics[width=1.0\textwidth]{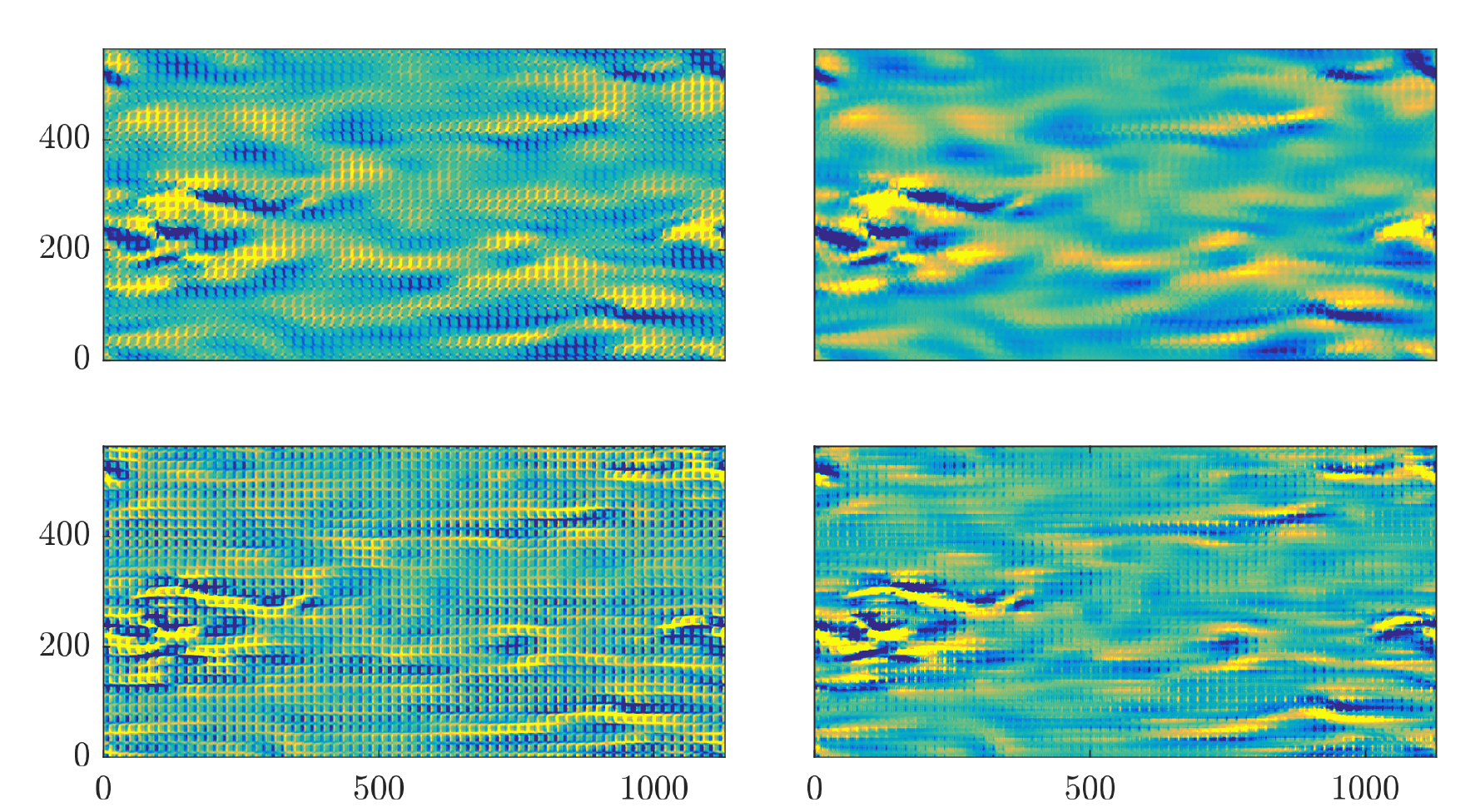}%
\mylab{-0.67\textwidth}{-.20cm}{$\Lxp$}%
\mylab{-0.19\textwidth}{-.20cm}{$\Lxp$}%
\mylab{-1.005\textwidth}{5.60cm}{$\Lzp$}%
\mylab{-1.005\textwidth}{1.94cm}{$\Lzp$}%
\mylab{-0.980\textwidth}{7.00cm}{\aaa.1)}%
\mylab{-0.980\textwidth}{3.35cm}{\bbb.1)}%
\mylab{-0.495\textwidth}{7.00cm}{\aaa.2)}%
\mylab{-0.495\textwidth}{3.35cm}{\bbb.2)}%
\vspace{0.1cm}%
\caption{Instantaneous realisation for case C09. (\aaa) spanwise velocity, at $y^+ \approx 0.3$. (\bbb) wall-normal velocity at $y^+ \approx 1$. (\aaa.1) and (\bbb.1), full signals; (\aaa.2) and (\bbb.2), turbulent contributions obtained using equations~\eqref{eq:flow_decomposition}.}
\label{fig:snapshots_decomposition_h09_bw} 
\end{figure}

The instantaneous overlying flow induces around roughness elements a velocity field with components in all three directions.
At the same time, the background turbulent flow has components in all three directions, so for instance the spanwise background shear induces a texture-coherent cross-flow.
Hence, the three velocity components have a roughness contribution induced by all components of the overlying flow.
Let us take the streamwise velocity signal as an example.
It would have contributions from the mean streamwise velocity, \meanU, the background-turbulent streamwise component, \uT, a roughness-coherent component, \uRu, driven and modulated by $\meanU + \uT$, and a roughness-coherent component, \uRw, driven and modulated by the background-turbulent spanwise component, \wT.
The simplified model presented in equation~\eqref{eq:decomposition_first_step} can then be extended to all components of the full velocity,
\begin{subequations}
  \begin{alignat}{5}
    u & \approx \meanU && + & \uT & + \frac{\meanU + \uT}{\meanU} \uRu && + \frac{\wT}{\tilW} \uRw && + \frac{\vT}{\tilV} \uRv, \label{eq:flow_decomposition_u}\\
    w & \approx        &&   & \wT & + \frac{\meanU + \uT}{\meanU} \wRu && + \frac{\wT}{\tilW} \wRw && + \frac{\vT}{\tilV} \wRv, \label{eq:flow_decomposition_w}\\
    v & \approx        &&   & \vT & + \frac{\meanU + \uT}{\meanU} \vRu && + \frac{\wT}{\tilW} \vRw && + \frac{\vT}{\tilV} \vRv,\label{eq:flow_decomposition_v}%
  \end{alignat}%
  \label{eq:flow_decomposition}%
\end{subequations}
where \uRw\ is the roughness-coherent streamwise velocity induced by the overlying, background spanwise shear of \wT, etcetera.
Note that, while the modulating signal $\meanU + \uT$ is always positive, \wT\ and \vT\ can either be positive or negative, and due to symmetry, their mean values cancel out.
To obtain a measure of the intensity of these components, \tilW\ and \tilV\ denote the conditional averages of $w$ and $v$, which account for their mean direction over individual roughness elements, such that they are only included into the average when they are positive \citep{GarM2011}.
In the cube roughness geometries studied in this work we observe no significant flow induced by $v$ in the streamwise and spanwise directions, although the corresponding terms have been included in equations~\eqref{eq:flow_decomposition} for completeness.
In textures with inclined planes, such as pyramids or cones, for instance, we would expect these contributions, \uRv\ and \wRv, to be relevant.
Figures~\ref{fig:snapshots_decomposition_u_h09_bw} and~\ref{fig:snapshots_decomposition_h09_bw} show instances of all three velocity components and their corresponding turbulent contributions, obtained by removing the roughness-coherent terms according to equations~\eqref{eq:flow_decomposition}.
The present decomposition, accounting for the modulation of the overlying turbulence, reduces significantly the signature of the roughness-coherent flow.
The remaining signature is overall much weaker than in the full signal, and is mostly prevalent in \vT.
For \uT, our results show essentially no footprint of the texture, even for the larger \kp\ studied.
In any event, the decomposition is based on a fundamentally linearised approach, assuming that the roughness lengthscales are much smaller than those over which the background overlying shear varies \citep{Zhang2016}, and can be expected to fail eventually as \kp\ is increase.

The modulation of the roughness-coherent flow by the background turbulence is analogous to the modulation of near-wall turbulence by the outer-layer dynamics \citep{Mathis2009}.
\citet{Zhang2016} have recently revisited this problem, and propose a formulation similar to equations~\eqref{eq:flow_decomposition}.
Also recently, \citet{Anderson2016} observed that the modulation of near-wall turbulence is present in flows over rough surfaces. 
In our direct numerical simulations, \Retau\ is not sufficiently high to reproduce large-scale dynamics, so the background turbulence is essentially that of near-wall, small-scale dynamics.
At larger \Retau, we could expect a cascading modulation, where large scales modulate small scales in the background turbulence, and the full background turbulence modulates the texture-coherent flow.

\begin{figure}
  \centering
  \vspace{.5cm}
  \includegraphics[width=\textwidth]{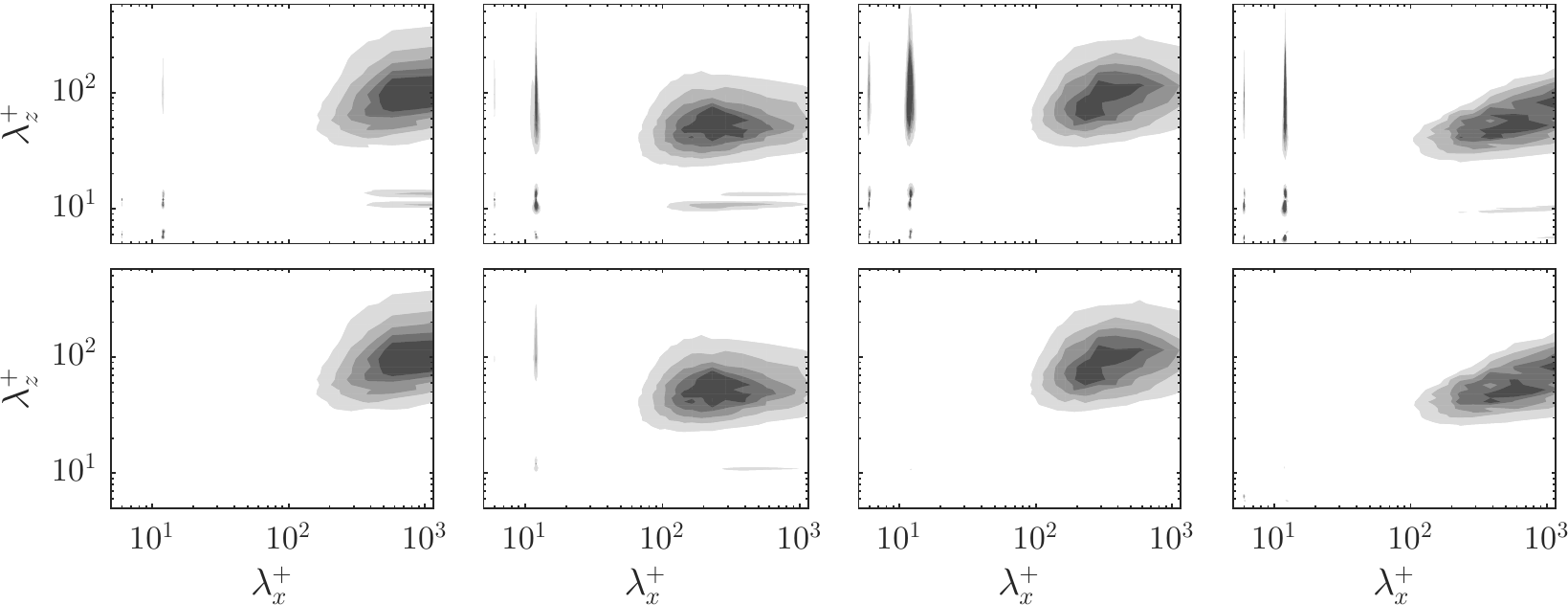}%
  \mylab{-0.865\textwidth}{5.35cm}{$ k_x k_z E^+_{uu}$}%
  \mylab{-0.625\textwidth}{5.35cm}{$ k_x k_z E^+_{vv}$}%
  \mylab{-0.390\textwidth}{5.35cm}{$ k_x k_z E^+_{ww}$}%
  \mylab{-0.165\textwidth}{5.35cm}{$-k_x k_z E^+_{uv}$}%
  \mylab{-1.040\textwidth}{4.1cm}{\rotatebox{90}{full}}%
  \mylab{-1.040\textwidth}{1.3cm}{\rotatebox{90}{turbulent}}%
  \caption{Premultiplied energy spectra of the full velocity signal and its background turbulent component, $k_x k_z E^+_{uu}$, $k_x k_z E^+_{vv}$ and $k_x k_z E^+_{ww}$, and cospectrum of the Reynolds stress, $k_x k_z E^+_{uv}$, for case C06, at $\yp = 2$ for $v$ and $\yp = 0.5$ for the rest.} 
  \label{fig:spectra_u_w_uv_clean_comparison_h06_y05}
  \centering
  \vspace{1.0cm}
  \includegraphics[width=\textwidth]{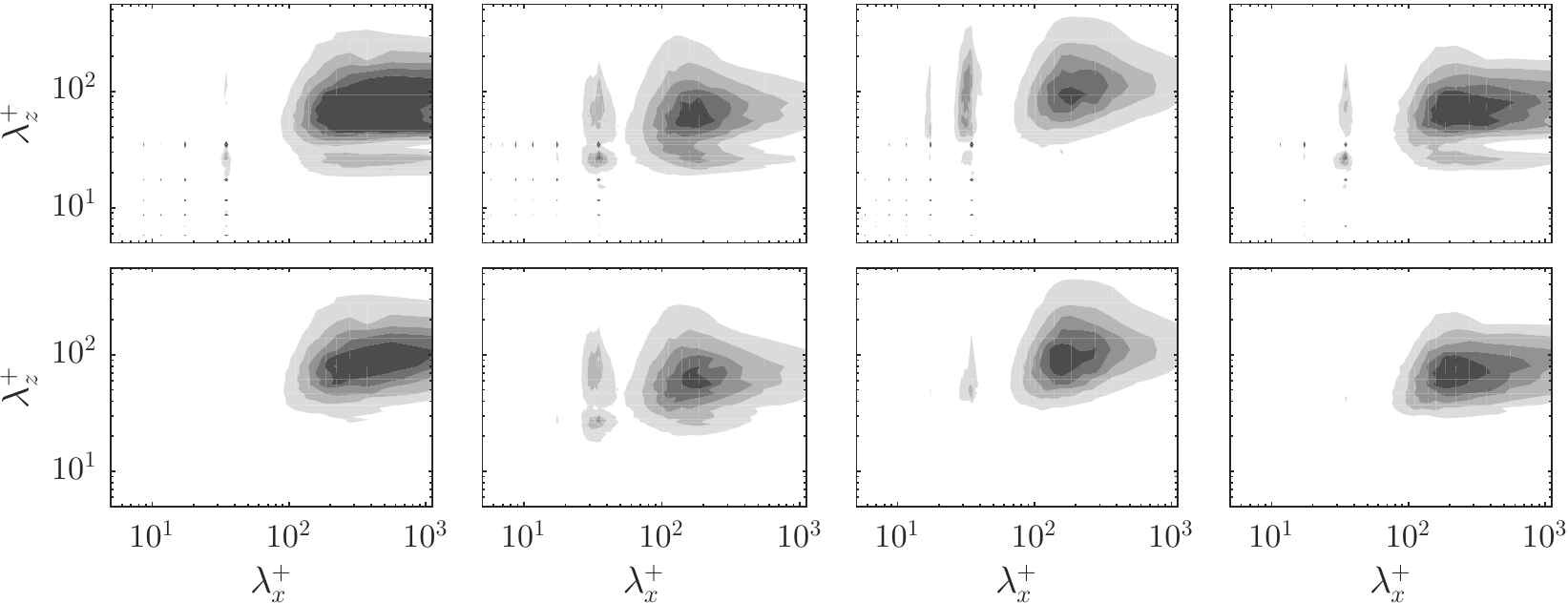}%
  \mylab{-0.865\textwidth}{5.35cm}{$ k_x k_z E^+_{uu}$}%
  \mylab{-0.625\textwidth}{5.35cm}{$ k_x k_z E^+_{vv}$}%
  \mylab{-0.390\textwidth}{5.35cm}{$ k_x k_z E^+_{ww}$}%
  \mylab{-0.165\textwidth}{5.35cm}{$-k_x k_z E^+_{uv}$}%
  \mylab{-1.040\textwidth}{4.1cm}{\rotatebox{90}{full}}%
  \mylab{-1.040\textwidth}{1.3cm}{\rotatebox{90}{turbulent}}%
  \caption{Premultiplied energy spectra of the full velocity signal and its background turbulent component, $k_x k_z E^+_{uu}$, $k_x k_z E^+_{vv}$ and $k_x k_z E^+_{ww}$, and cospectrum of the Reynolds stress, $k_x k_z E^+_{uv}$, for case C18, at $\yp = 2$ for $v$ and $\yp = 0.5$ for the rest.} 
  \label{fig:spectra_u_w_uv_clean_comparison_h18_y05}
\end{figure}
Beyond instantaneous realisations, the spectral densities of the different flow variables provide quantitative, statistical evidence to evaluate the proposed decomposition.
Figures~\ref{fig:spectra_u_w_uv_clean_comparison_h06_y05} and~\ref{fig:spectra_u_w_uv_clean_comparison_h18_y05} show premultiplied spectra and cospectra of the full velocity, as well as those of the background turbulence variables, for a roughness of $\kp \approx 6$ and one of $\kp \approx 18$.
The full flow signals exhibit a large spike at the wavelengths of the texture, \sx\ and \sz, as well as its corresponding subharmonics.
However, the surrounding wavelengths also contain a substantial amount of energy, in a similar fashion to the effect observed in figure~\ref{fig:one_d_cut_phys_and_spect_u_h06_bw_JFM} in one dimension.
These regions are the signature of the amplitude modulation of the roughness-coherent flow.
Using the classical triple decomposition, without the modulation of the roughness-coherent component, removes the wavelength of the texture only, as shown in figure~\ref{fig:one_d_cut_phys_and_spect_u_h06_bw_JFM}(\bbb.2).
In addition to the wavelengths of the roughness-coherent flow and the background turbulence, we observe very elongated regions with the wavelength of the roughness in $z$ and the range of wavelengths of the background turbulence in $x$, or vice versa.
These are the signature of cross-terms BT--RC in equation~\eqref{eq:flow_decomposition}.
Let us take for instance the term \uT\uRu, which essentially arises from the interaction of overlying streaks with the texture.
In figure~\ref{fig:snapshots_decomposition_u_h09_bw}(\aaa), it can be observed that the region of high $u$ corresponding to a streak is broken down by the canyons formed by the roughness, so that within the footprint of the streak there are streamwise-aligned, alternating stripes of high and low $u$.
The signature of this in spectral space will have the streamwise wavelength of streaks, but the spanwise wavelength of the texture.

For small roughness textures, as in figure~\ref{fig:spectra_u_w_uv_clean_comparison_h06_y05}, the scale separation between the turbulent and roughness contributions is large enough that they can essentially be extracted by Fourier filtering.
We exploited this in the example shown in figures~\ref{fig:one_d_cut_phys_and_spect_u_h06_bw_JFM}(\ccc.1) and~\ref{fig:one_d_cut_phys_and_spect_u_h06_bw_JFM}(\ccc.2).
However, for larger \kp, as shown in figure~\ref{fig:spectra_u_w_uv_clean_comparison_h18_y05}, the elongated regions mentioned above overlap with the turbulent contribution, which makes a pass-band filter ill-suited.
Nevertheless, equations~\eqref{eq:flow_decomposition}, interpreted as a filter, provide a tool to extract the turbulent contribution when Fourier filtering is not possible due to an overlap of wavelengths.
In \S\ref{sec:the_turbulent_component} this decomposition is used to extract the background turbulence component and analyse the effects of roughness on it.
Before that, in \S\ref{sec:rough_comp}, the roughness-coherent contribution is characterised, and a model to predict it is proposed.

\section{The roughness-coherent component}\label{sec:rough_comp}

In smooth walls, the mean velocity is a function of the wall-normal coordinate alone.
However, in rough surfaces, we can define a time-averaged mean velocity that not only depends on $y$, but is also a function of the $x$ and $z$ coordinates.
This space-dependent mean flow minus the conventional mean velocity, $U(y)$, is the roughness-coherent contribution introduced in the previous section, which is steady and coherent in space with the roughness texture. 
If we focus on individual roughness elements, the coherent flow can be thought of as the flow around the obstacles of the roughness geometry driven by the overlying shear.
As the flow is deflected from and surrounds the obstacles, it generates a three-dimensional velocity field.
The roughness-coherent component can be obtained by averaging the flow in time.
Since all geometries in the present work are made up of a periodic pattern, the roughness-coherent contribution will also be periodic, with the same wavelength of the roughness texture.
This contribution is therefore obtained by ensemble averaging over time and over the periodic texture units as in \citet{Seo2015}.

\begin{figure}
  \centering
  \includegraphics[width=1.0\textwidth]{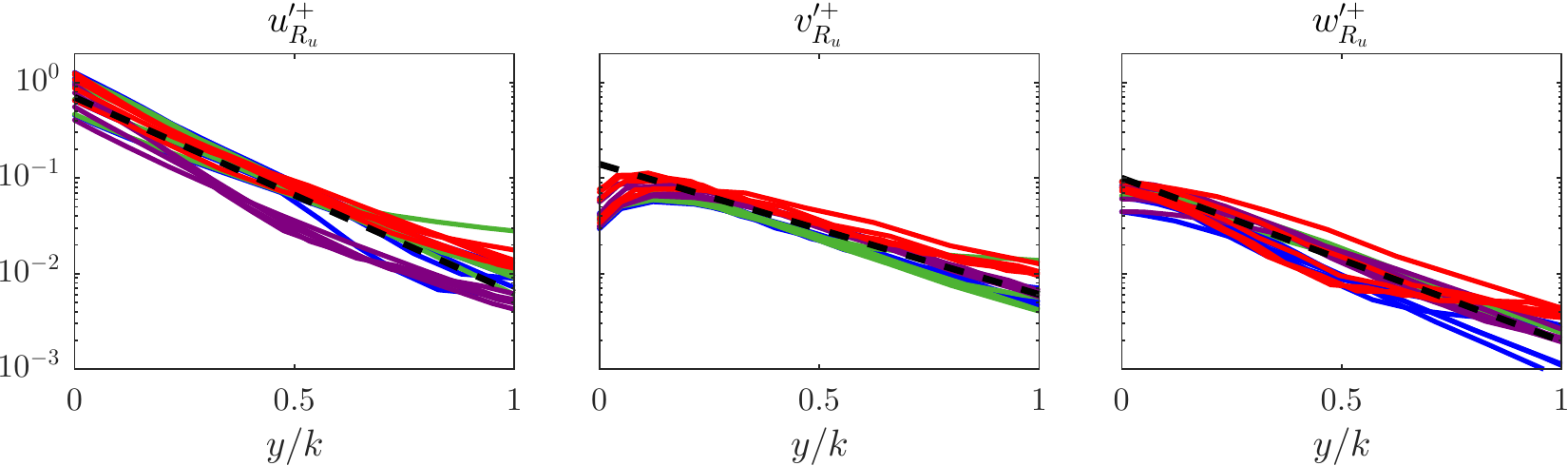}
  \caption{Rms of the roughness-coherent velocity components above the roughness crests. \protect\blueline, collocated posts; \protect\greenline, spanwise-staggered posts; \protect\purpleline, streamwise-staggered posts; \protect\redline, collocated posts of two heights. The reference lines, \protect\blacklinedash, have slopes of $1.5\pi$, $1\pi$ and $1.25\pi$, respectively.}
  \label{fig:exponential_decay_rough_component}
\end{figure}
The coherent flow thus obtained decays above the roughness crests with height.
In all our cases, the roughness-coherent components of the velocity have heavily decayed at a distance of $\sim\k$, shorter than the commonly accepted height of the roughness-sublayer, $\sim3\k$ \citep{Raupach1991,Flack2007}.
The rms of the roughness-coherent components of the velocity, whose squares are commonly referred to as dispersive stresses \citep{Raupach1982}, are depicted in figure~\ref{fig:exponential_decay_rough_component} for several roughness textures and sizes.
The decay is of the form $\sim \exp(-y/k)$ and is observed to scale with the roughness height, \k, rather than the spacing, $s$.
The rate of decay is observed to be different for the three velocities, and is fastest for $u$ and slowest for $v$.
For all direct numerical simulations, the magnitude of the fluctuations, and in particular that of the wall-normal component, is below $2$ percent of \utau\ at one roughness height above the roughness crests.
These results are consistent with experiments, where dispersive stresses are found to vanish at $y \approx \k$ \citep{Cheng2002,Florens2013}.

For vanishingly small \k, the roughness-coherent flow can be thought of as the flow induced around the roughness elements driven by a steady, homogeneous overlying shear.
This concept was already used by \citet{Luc1991}, where they considered riblets of a vanishingly small size, so that they perceive the overlying turbulence as steady and homogeneous.
Notice that, for small roughness, in the transitionally rough regime, the time-scales and length-scales of the overlying background-turbulent fluctuations are much larger than those of roughness.
These latter fluctuations are quasi-steady and quasi-homogeneous with respect to the characteristic scales of the surface, i.e.\ they vary slowly and over long distances.
The concept of a quasi-steady, quasi-homogeneous limit was formalised by \citet{Zhang2016} for the interaction between the overlying large-scales of the outer region and the smaller and faster near-wall fluctuations, and a similar framework would apply here.

\begin{figure}
  \centering
  \includegraphics[width=1\textwidth]{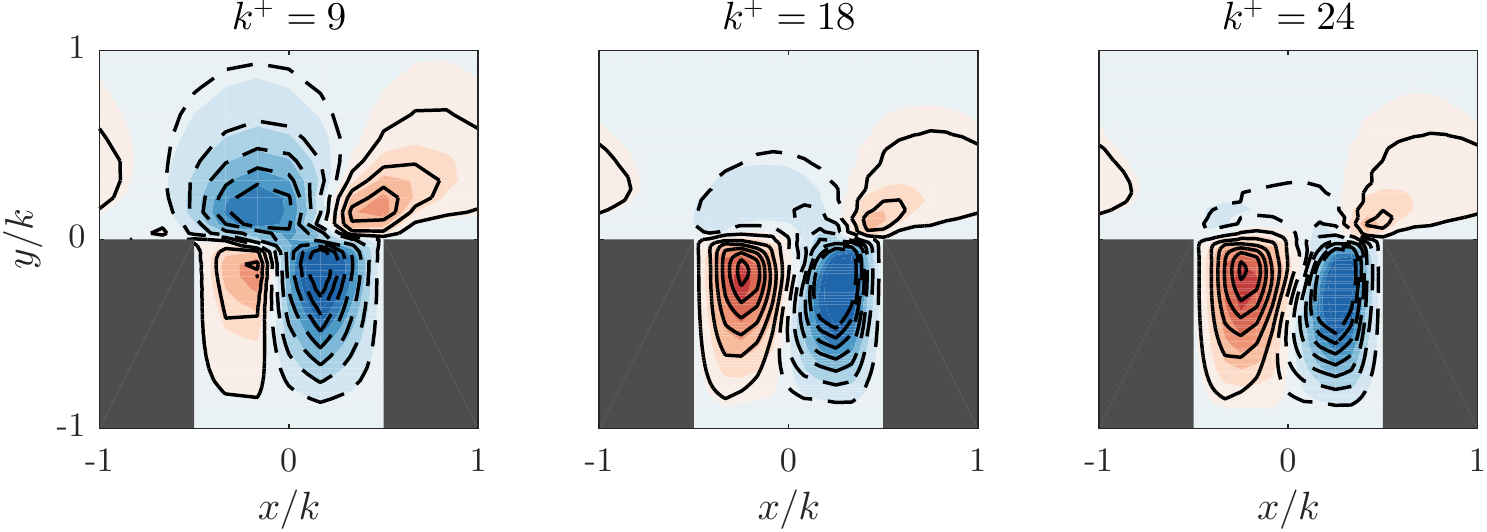}
  \caption{Wall-normal velocity for cases (left) C09, (centre) C18 and (right) C24. Shaded, roughness-coherent contribution from direct numerical simulations; contours, steady laminar model. Red and solid indicate positive values, blue and dashed indicate negative values.}
  \label{fig:vR_vertical_09_18_24}
\end{figure}
\begin{figure}
  \centering
  \includegraphics[width=1.0\textwidth]{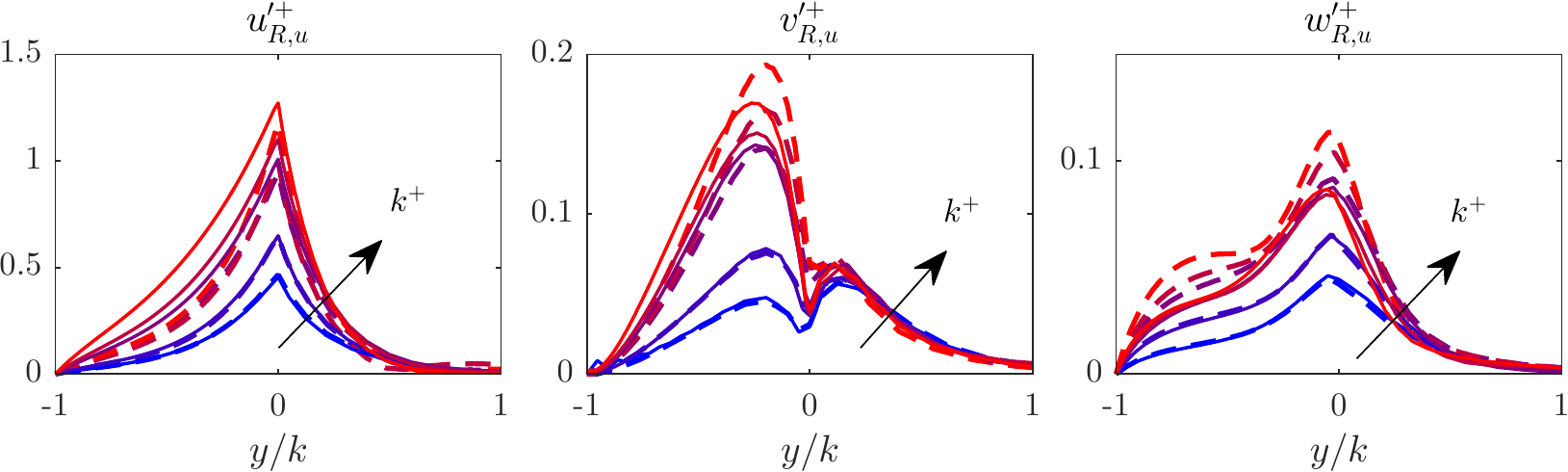}
  \includegraphics[width=1.0\textwidth]{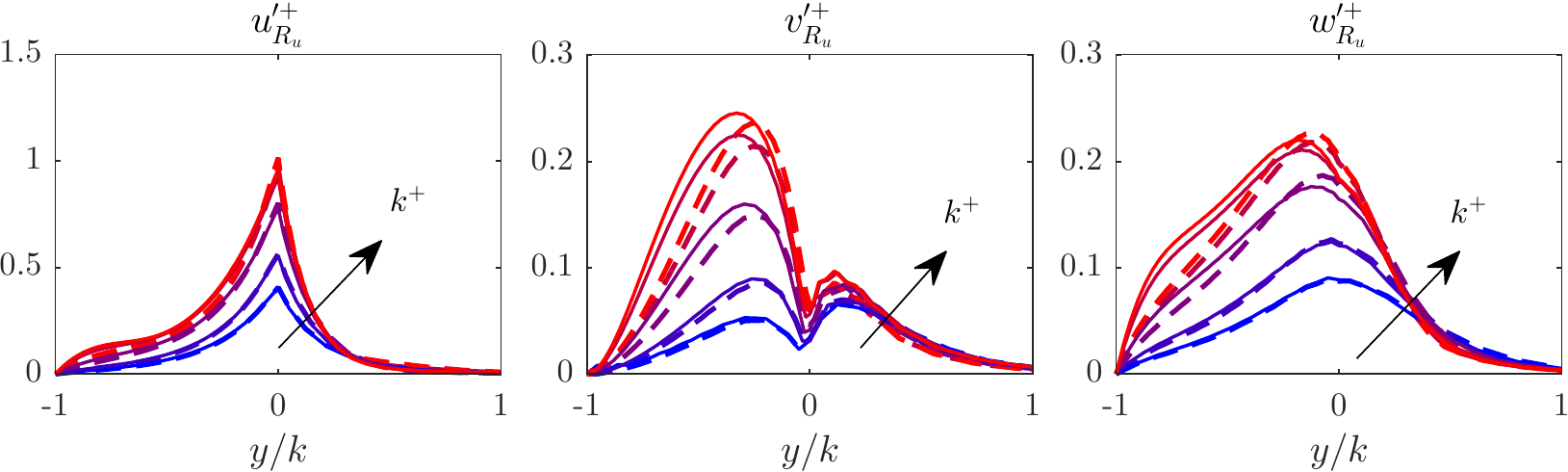}
  \caption{Rms of the roughness-coherent contribution for the collocated and streamwise-staggered roughness. Solid, results from direct numerical simulations; dashed, results from the laminar model simulations. Blue to red, cases C06 to C18 and SX06 to SX18.}
  \label{fig:rms_comparison_laminar_1}
\end{figure}
\begin{figure}
  \centering
  \includegraphics[width=.75\textwidth]{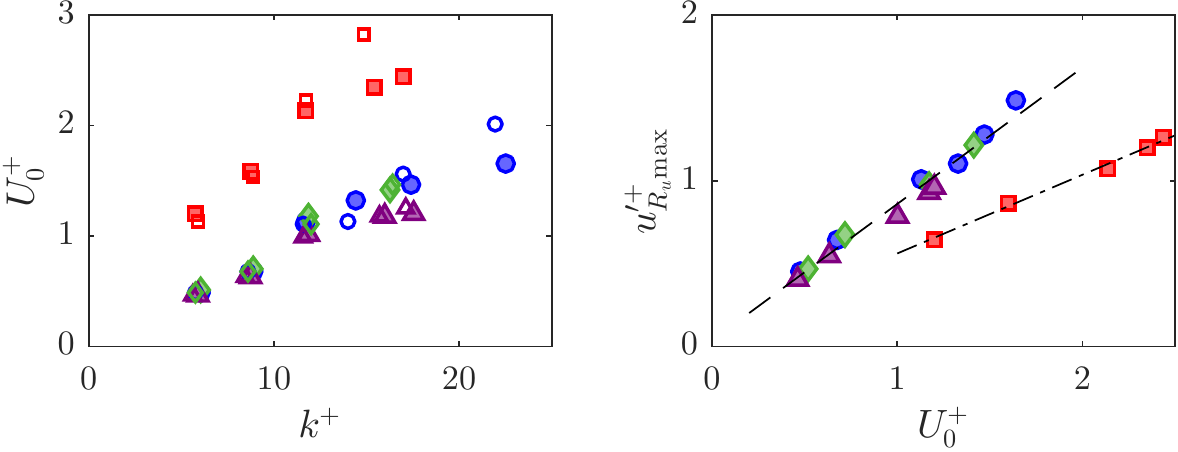}%
  \mylab{-0.682\textwidth}{3.40cm}{\aaa)}%
  \mylab{-0.285\textwidth}{3.40cm}{\bbb)}%
  \caption{(\aaa) Mean streamwise velocity at the roughness crests. (\bbb) Maximum streamwise roughness-coherent velocity as a function of the mean streamwise velocity at the roughness tips. \protect\bluecircle, collocated posts; \protect\greendiamond, spanwise staggered posts; \protect\purpletriangle, streamwise-staggered posts; \protect\redsquare, collocated posts of two heights. Empty symbols correspond to estimations from the laminar model. Linear regression with slopes: \dashed, $0.82$; and \chndot, $0.47$.}
  \label{fig:velocity_tips}
\end{figure}
The roughness-coherent flow for small \kp, as suggested above, can be modelled as the flow induced by a steady and homogeneous overlying background flow.
Therefore, to approximate the roughness-coherent contribution, we conduct steady, laminar simulations using the numerical methodology described in \S\ref{sec:methodology}, but where the periodic domain only contains one texture element.
The mean shear and viscosity are adjusted to match the same \kp\ of the corresponding direct numerical simulations.
For small but finite values of \kp, these numerical domains are too small to sustain turbulence \citep{Jimenez1991}, and result in laminar, steady flows, which provide estimates of the roughness-coherent contribution.
For instance, figure~\ref{fig:vR_vertical_09_18_24} compares the wall-normal roughness-coherent velocity obtained from the direct numerical simulations by ensemble averaging to the laminar model.
Laminar simulations begin to deviate for the intermediate case, C18, but they still exhibit good qualitative agreement for the larger case, C24.
Notice that, even for the smallest \kp, the flow is not symmetric and does not behave as purely viscous. 
It is therefore necessary to consider the advective terms, in contrast with the Stokes-flow analysis of \citet{Luc1991}.
This model also allows us to predict the rms fluctuations of the roughness-coherent flow. 
Figure~\ref{fig:rms_comparison_laminar_1} depicts data from our direct numerical simulations compared against those from the laminar model, and shows good agreement for $\kp \lesssim 15$ for all our roughness surfaces.
Notice that this method can also be used to estimate properties of the mean velocity profile close to the rough surface, such as the mean streamwise velocity at the roughness crests, by averaging along the $x$ and $z$ directions, as shown in figure~\ref{fig:velocity_tips}.

\section{The background turbulence component}\label{sec:the_turbulent_component}

\begin{figure}
  \centering
  \includegraphics[width=1.0\textwidth]{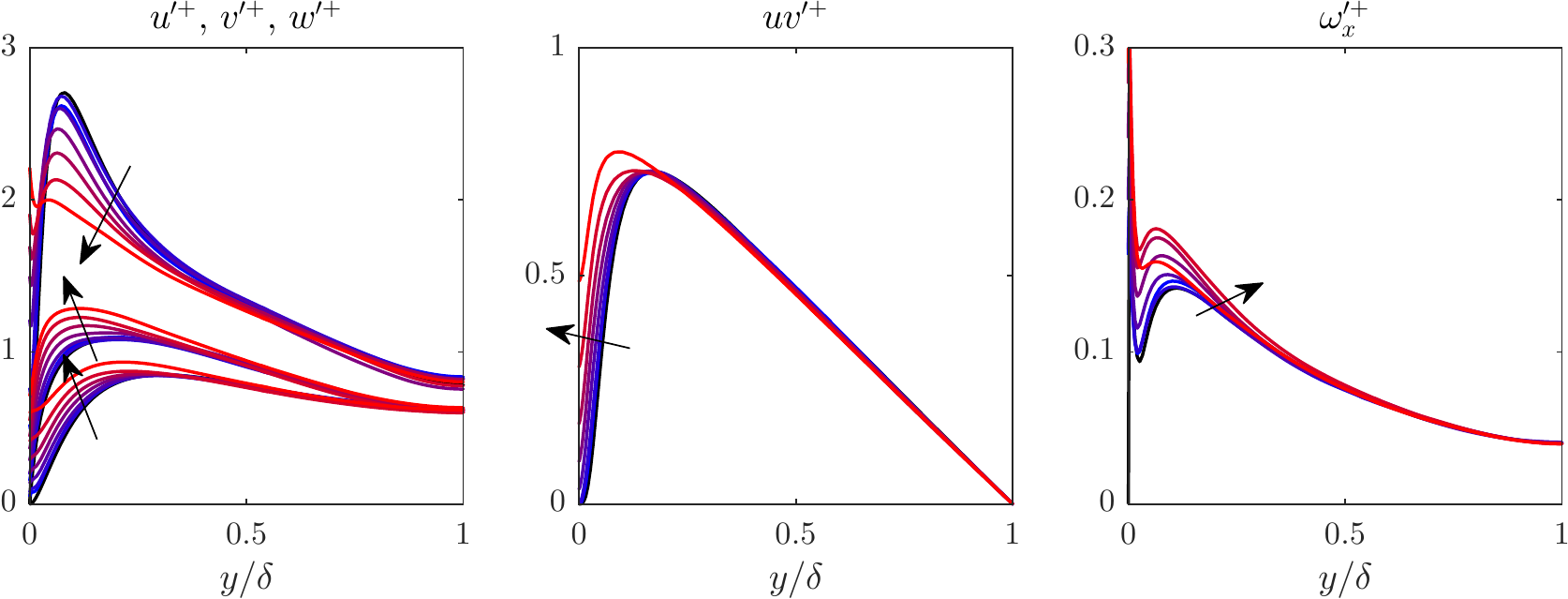}%
  \mylab{-0.915\textwidth}{3.40cm}{$u^{\prime +}$}%
  \mylab{-0.915\textwidth}{1.50cm}{$v^{\prime +}$}%
  \mylab{-0.940\textwidth}{2.56cm}{$w^{\prime +}$}%
  \mylab{-0.732\textwidth}{4.4cm}{\aaa)}%
  \mylab{-0.382\textwidth}{4.4cm}{\bbb)}%
  \mylab{-0.032\textwidth}{4.4cm}{\ccc)}%
  \caption{Full rms fluctuations. Blue to red, cases C06 to C36; black, smooth channel. The arrows indicate increasing roughness size. (\aaa) Streamwise, wall-normal and spanwise velocity fluctuations; (\bbb) Reynolds stress; (\ccc) streamwise vorticity fluctuations.}
  \label{fig:rms_outer_units_uvw}
\end{figure}

Roughness does not only excite the wavelengths of the surface texture, but also modifies the background turbulence.
This can be observed in the rms fluctuations that incorporate effects from both the roughness contribution, studied in the previous section, and the background turbulence.
The streamwise rms fluctuations, shown in figure~\ref{fig:rms_outer_units_uvw}, decrease near the wall, and in particular the peak of intensity lowers as the roughness size increases, while a significant growth occurs at the roughness crests in a similar fashion to that in figure~\ref{fig:rms_comparison_laminar_1}.
This decrease in the peak of intensity can only be caused by a decrease, and therefore modification, of the rms fluctuations of the background-turbulent component.
The wall-normal and spanwise rms fluctuations present an increase of intensity across the entire roughness-sublayer, from the roughness crests to their near-wall intensity peak.
The near-wall intensity peak of the streamwise vorticity fluctuation also increases, save for the largest case, which flattens in magnitude and almost disappears, this being a symptom of roughness altering the near-wall cycle.
Notice that these modifications of the rms fluctuations extend to a distance of $\sim 2\k$, larger than the decay distance of the roughness-coherent signal observed in the previous section.
However, beyond that distance, the rms fluctuations resemble those in smooth-wall turbulence, in agreement with the outer-layer similarity hypothesis \citep{Townsend1976}.

In this section, we analyse the effect of roughness on the background turbulence and, in particular, on the rms fluctuations.
The rms fluctuations of the full signal, as shown in figure~\ref{fig:rms_outer_units_uvw}, contain contributions from both the turbulent-background and roughness-coherent components.
However, the decomposition of equations~\eqref{eq:flow_decomposition} can be used to derive approximate expressions for the rms velocities from which the rms fluctuations of the background turbulence can be extracted.
For the geometries studied in this paper, the leading terms are
\begin{subequations}
  \begin{align}
  \langle u^{\prime 2} \rangle 
  =& \left< \uT^{2} \right> 
  +  \left< \uRu^2 \right> \\
  +& \left< \uRu^{2} \right> \left< \frac{\uT^{2}}{\meanU^2} \right> 
  +  \left< \uRv^{2} \right> \left< \frac{\vT^{2}}{\tilV^2} \right>
  +  \left< \uRw^{2} \right> \left< \frac{\wT^{2}}{\tilW^2} \right>, \nonumber \\
  \langle v^{\prime 2} \rangle 
  =& \left< \vT^{2} \right> 
  +  \left< \vRu^2 \right> \\
  +& \left< \vRu^{2} \right> \left< \frac{\uT^{2}}{\meanU^2} \right> 
  +  \left< \vRv^{2} \right> \left< \frac{\vT^{2}}{\tilV^2} \right>
  +  \left< \vRw^{2} \right> \left< \frac{\wT^{2}}{\tilW^2} \right>, \nonumber \\
  \langle w^{\prime 2} \rangle 
  =& \left< \wT^{2} \right> 
  +  \left< \wRu^2 \right> \\
  +& \left< \wRu^{2} \right> \left< \frac{\uT^{2}}{\meanU^2} \right> 
  +  \left< \wRv^{2} \right> \left< \frac{\vT^{2}}{\tilV^2} \right>
  +  \left< \wRw^{2} \right> \left< \frac{\wT^{2}}{\tilW^2} \right>, \nonumber \\
  \langle u^{\prime} v^{\prime} \rangle 
  =& \left< \uT \vT \right> 
  +  \left< \uRu \vRu \right> \\
  +& \left< \uRu \vRu \right> \left< \frac{\uT^{2}}{\meanU^2} \right> 
  +  \left< \uRv \vRv \right> \left< \frac{\vT^{2}}{\tilV^2} \right> 
  +  \left< \uRw \vRw \right> \left< \frac{\wT^{2}}{\tilW^2} \right> \nonumber \\
  +& \left< \uRu \vRv \right> \left< \frac{\uT \vT}{\meanU \tilV} \right>,\nonumber
  \end{align}%
  \label{eq:flow_decomposition_rms}%
\end{subequations}%
where the angled brackets indicate temporal and x-z-spatial averaging, and the prime refers to fluctuations with respect to the mean.
The full expressions including the terms that are negligible for our geometries can be found in appendix~\ref{appendixA}. 
Equations~\ref{eq:flow_decomposition_rms} are adequate approximations only for small roughness, as long as equations~\eqref{eq:flow_decomposition} hold.
The coherent--background cross terms, which are zero in the conventional triple decomposition, can be of the same order of magnitude as the coherent--coherent terms.
Equations~\eqref{eq:flow_decomposition_rms} can be used to extract the rms fluctuations of the background-turbulent contribution from the full and the roughness-coherent signals.

\begin{figure}
  \vspace{0.3cm}
  \centering
  \includegraphics[width=\textwidth]{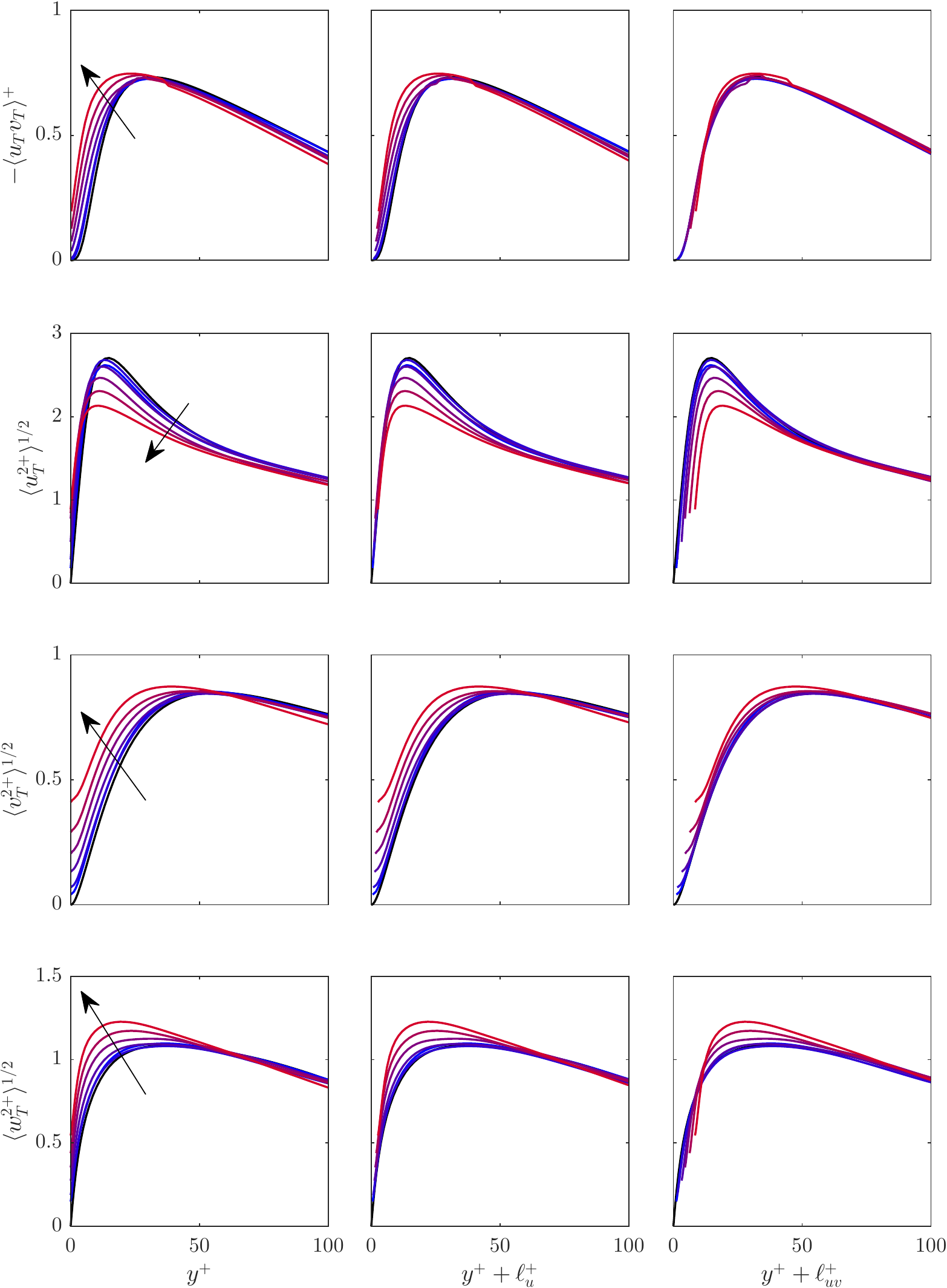}%
  \mylab{-0.700\textwidth}{14.85cm}{\aaa.1)}%
  \mylab{-0.383\textwidth}{14.85cm}{\aaa.2)}%
  \mylab{-0.063\textwidth}{14.85cm}{\aaa.3)}%
  \mylab{-0.700\textwidth}{10.25cm}{\bbb.1)}%
  \mylab{-0.383\textwidth}{10.25cm}{\bbb.2)}%
  \mylab{-0.063\textwidth}{10.25cm}{\bbb.3)}%
  \mylab{-0.700\textwidth}{05.67cm}{\ccc.1)}%
  \mylab{-0.383\textwidth}{05.67cm}{\ccc.2)}%
  \mylab{-0.063\textwidth}{05.67cm}{\ccc.3)}%
  \mylab{-0.700\textwidth}{01.10cm}{\ddd.1)}%
  \mylab{-0.383\textwidth}{01.10cm}{\ddd.2)}%
  \mylab{-0.063\textwidth}{01.10cm}{\ddd.3)}%
  \caption{Rms fluctuations of the background turbulent flow. (\aaa) Reynolds stress, \uT\vT; (\bbb) streamwise velocity, \uT; (\ccc) wall-normal velocity, \vT; (\ddd) spanwise velocity, \wT. Blue to red, cases C06 to C24; black, smooth channel; the arrow indicates increasing \kp.}
  \label{fig:rms_clean_uv}
\end{figure}
\begin{figure}
  \centering
  \hspace{-0.8cm}%
  \includegraphics{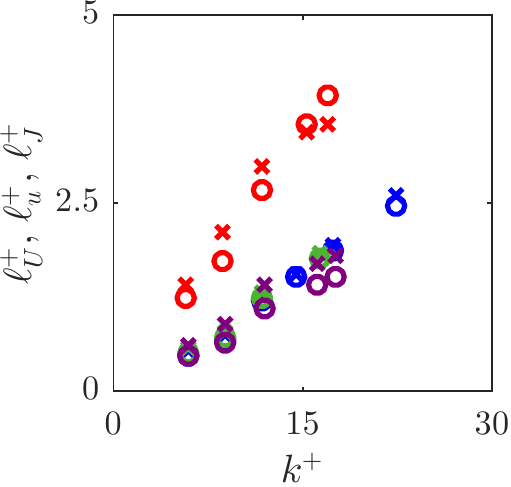}%
  \caption{Virtual origins for different geometries and \kp. \protect\blackcross, \protup; \protect\emptyblackcircle, \Dup; and \protect\blackplus, \Djp. Blue  collocated posts; green, spanwise-staggered posts; purple, streamwise-staggered posts; red, collocated posts of two heights.}
  \label{fig:protrusion_height_and_Du}
\end{figure}
The rms fluctuations of the background turbulence are shifted towards the wall, as if they perceived a smooth wall, or a virtual origin, below the roughness crests.
In the first column of panels in figure~\ref{fig:rms_clean_uv}, the rms fluctuations of the background turbulence are compared with those in smooth-wall turbulence.
The decomposition of equations~\eqref{eq:flow_decomposition_rms} removes from the background turbulence rms's the near-wall peaks observed for the full rms's in figure~\ref{fig:rms_outer_units_uvw}.
The fluctuations are shifted towards the wall, but otherwise display a similar shape to those of smooth-wall turbulence close to the wall.
We refer to this displacements as virtual origins, since the rms fluctuations behave as if they had an origin below the roughness crests.
This can be interpreted as the height below the roughness crests at which they would to go to zero if extended as smooth-wall rms fluctuations.

The virtual origin of the streamwise rms fluctuations, \Dup, is obtained as the depth below the roughness crests at which $\uT^{\prime +}$ would zero out, when extrapolated from its profile above the crests.
When $\uT^{\prime+}$ is portrayed versus the height measured from that origin, as in figure~\ref{fig:rms_clean_uv}(\bbb.2), a good collapse with smooth wall data is observed in the first few wall units of height.
However, this collapse does not extend outside the roughness-sublayer, where all curves should converge to the smooth-wall case.
According to Townsend's outer-layer similarity hypothesis \citep{Townsend1976}, sufficiently far from the wall, effectively outside the roughness sublayer, all turbulent fluctuations are independent of the surface condition when normalised in wall units.
In the second column of panels in figures~\ref{fig:rms_clean_uv}, the origin of the rms fluctuations is set at \Dup\ for all variables.
We observe that this virtual origin does not adequately collapse any rms fluctuation other than that of the streamwise velocity very near the wall.

We observe that the streamwise virtual origin, \Du, is related to the apparent origin of the mean velocity profile.
Determining the correct zero-plane for the law of the wall is a question that has always been present in roughness studies.
This reference level for the mean velocity profile is usually referred to as displacement height, and it is in essence equivalent to what we have defined as virtual origins, as it is a shift in the $y$-coordinate.
\citet{Jackson1981} proposes a displacement height, \Dj, based on the centroid of the total stress below the roughness crests, i.e.\ the height at which the mean surface drag appears to act,
\begin{equation}
  \Djp = \int_{\kp} \left(\frac{dU^{+}}{dy^{+}} {\ReStressp} \right) d\yp.
  \label{eq:JacksonDisplacementHeight}
\end{equation}
Similarly, since this displacement is essentially a virtual origin, we can also define another virtual origin for the mean velocity profile, $\protup$, as the distance from the roughness crests at which it would go to zero. 
This definition is equivalent to that used above for \Dup, where $\protup$ can be obtained by linearly extrapolating the mean velocity profile at the wall.
In figure~\ref{fig:protrusion_height_and_Du}, we observe that $\protup$, \Dup\ and \Djp\ are generally similar for different roughness configurations and sizes.

Further away from the wall, where the influence of the roughness-coherent flow is negligible, the rms fluctuations suggest that turbulence behaves as smooth-wall, canonical turbulence with a virtual origin \Duvp.
The virtual origin of the Reynolds shear stress, \Duv\, appears to be the virtual origin of turbulence outside the roughness sublayer \citep{GGMadrid2018,Fairhall2018}.
Unlike for the rms fluctuations of the three velocity components, the shape near the wall of the Reynolds shear stress does not drastically change with roughness.
In figure~\ref{fig:rms_clean_uv}(\aaa.3), we observe that using the virtual origin of its rms fluctuations, \Duv, not only collapses the region near the wall, but the entire curve.
In the third column of panels in figure~\ref{fig:rms_clean_uv}, the origin of the rms fluctuations is set at \Duv\ for all variables.
The rms fluctuations, including the streamwise ones, converge to smooth-wall turbulence rms fluctuations outside the roughness-sublayer.
Very near the wall, different variables would extrapolate to zero at different heights, as is particularly evident for $\uT^{\prime+}$ and $\wT^{\prime+}$, which experience virtual origins shallower than \Duvp.
Beyond this, up to at least $\kp \approx 15$, the curves show an excellent collapse with smooth wall data.
This suggests that, up to that \kp, turbulence remains essentially canonical, i.e.\ smooth-wall-like.

\begin{figure}
  \centering
  \vspace{.6cm}
  \includegraphics[width=\textwidth]{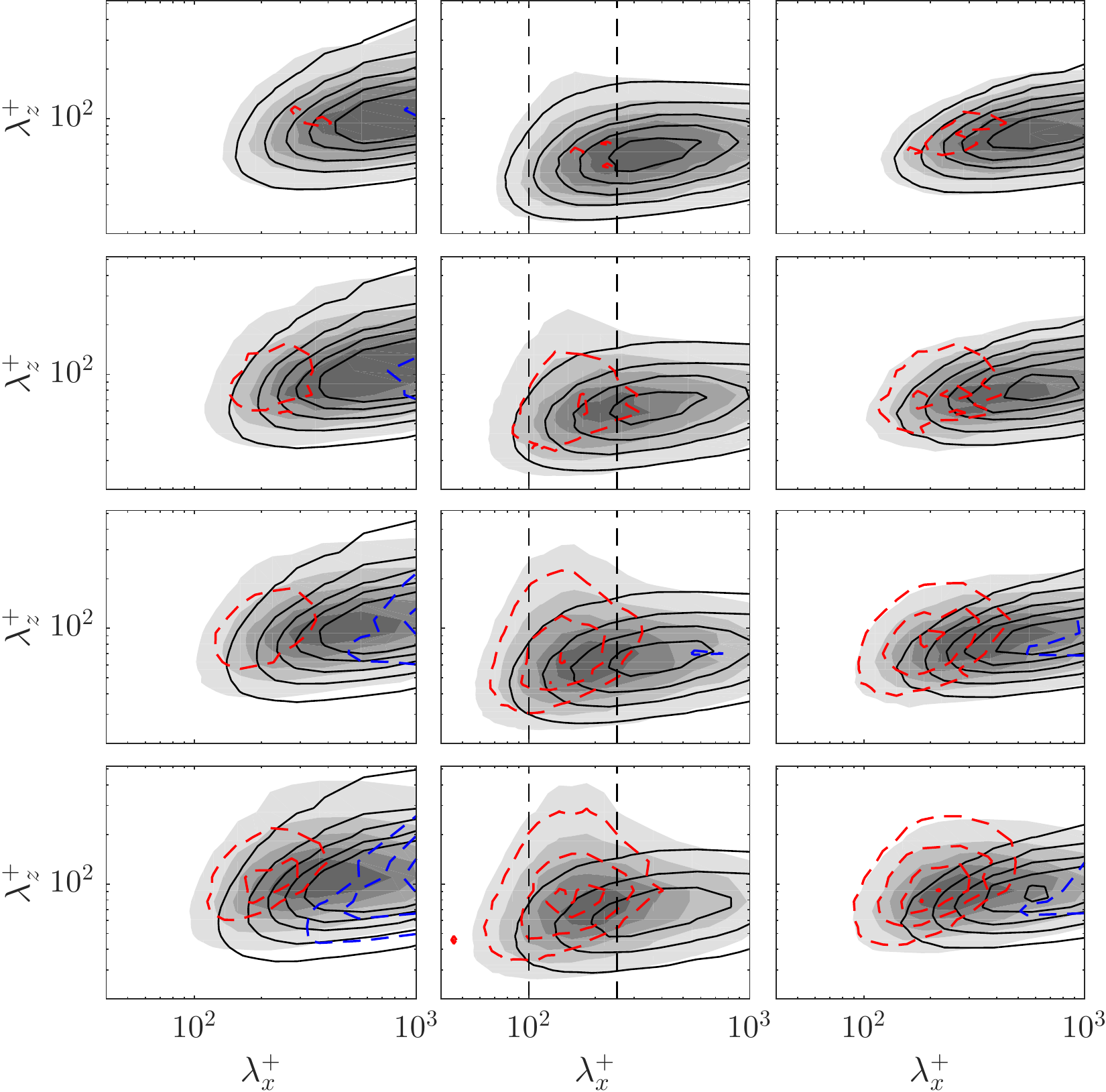}%
  \mylab{-0.82\textwidth}{13.6cm}{$   k_x k_z E_{uu}^+$}%
  \mylab{-0.51\textwidth}{13.6cm}{$   k_x k_z E_{vv}^+$}%
  \mylab{-0.21\textwidth}{13.6cm}{$ - k_x k_z E_{uv}^+$}%
  \caption{Two-dimensional premultiplied spectra of the streamwise and wall-normal velocity components, and cospectrum of the Reynolds stress at $\yp = 8$, from top to bottom cases C12, C15, C18 and C24. Shaded, background-turbulent component; \protect\blackline, smooth-wall data at the equivalent height above the corresponding virtual origin; \protect\redlinedash\ and \protect\bluelinedash, positive and negative difference between the rough and smooth-wall results. Vertical dashed lines indicate $\lambda_x^+ = 100$ and $250$. Contours indicate the same value in wall units.}
  \label{fig:outer_spectra_u_v_uv_y8}
\end{figure}
To further explore the modifications of the background turbulence by roughness within the roughness-sublayer, we analyse its two-dimensional energy spectra and cospectra.
As introduced in \S\ref{sec:flow_decomp}, the decomposition can be used to obtain the two-dimensional energy spectra of the background turbulence, without the footprint of roughness.
Figure~\ref{fig:outer_spectra_u_v_uv_y8} shows the turbulent premultiplied spectra of \uT, \vT\ and premultiplied cospectra of \uT\vT, together with smooth-wall results at the equivalent height accounting for the corresponding virtual origin.
For this roughness, the spectra present little changes for $\kp \lesssim 12$.
As size increases the spectra and cospectra start to be modified: the energy at large $\lambda_x^+$ decreases, while at smaller $\lambda_x^+$ there is an increase of energy.
The decrease of energy at large wavelengths is particularly clear for the streamwise velocity component.
There is also a noticeable change in the spanwise direction.
The spectra and cospectra, especially that of $v$, display an increase of energy at small $\lambda_x^+$ and a wider range of $\lambda_z^+$.
This increase of energy is centred approximately at $\lambda_x^+ \approx 150$, being particularly clear for $E_{vv}$.
\section{Shear-flow instability}\label{sec:kelvin_helmholtz}

The spectral energy densities in figure~\ref{fig:outer_spectra_u_v_uv_y8} indicate the regions where energy increases or decreases with respect to smooth wall turbulence.
Energy increases at short streamwise wavelengths, $\lambda_x$, for a broad range of $\lambda_z$, both larger and smaller than the characteristic $\lambda_z$ of smooth-wall turbulence.
The values of $\lambda_x$ at which energy increases, $\lambda_x^+ \approx 100$--$200$, appear to be independent of \kp.
Although significantly more intense, similar modifications of the energy spectra were observed on riblets, a particular case of roughness. 
This increase in energy was found to be due to the formation of a shear flow instability \citep{GarM2011}.
These instabilities are a common feature in obstructed flows \citep{Ghisalberti2009}, and have been observed on flows over plant canopies \citep{Fin2000} and permeable substrates \citep{Bre2006}.
Although we do not observed them directly in our numerical simulations, the concentration of energy for a narrow range of $\lambda_x^+$ suggests a receptivity to this wavelengths that could be connected to a similar shear-flow instability of the mean velocity profile, for which we analyse the stability properties in this section.

\begin{figure}
\centering
\includegraphics[width=\textwidth]{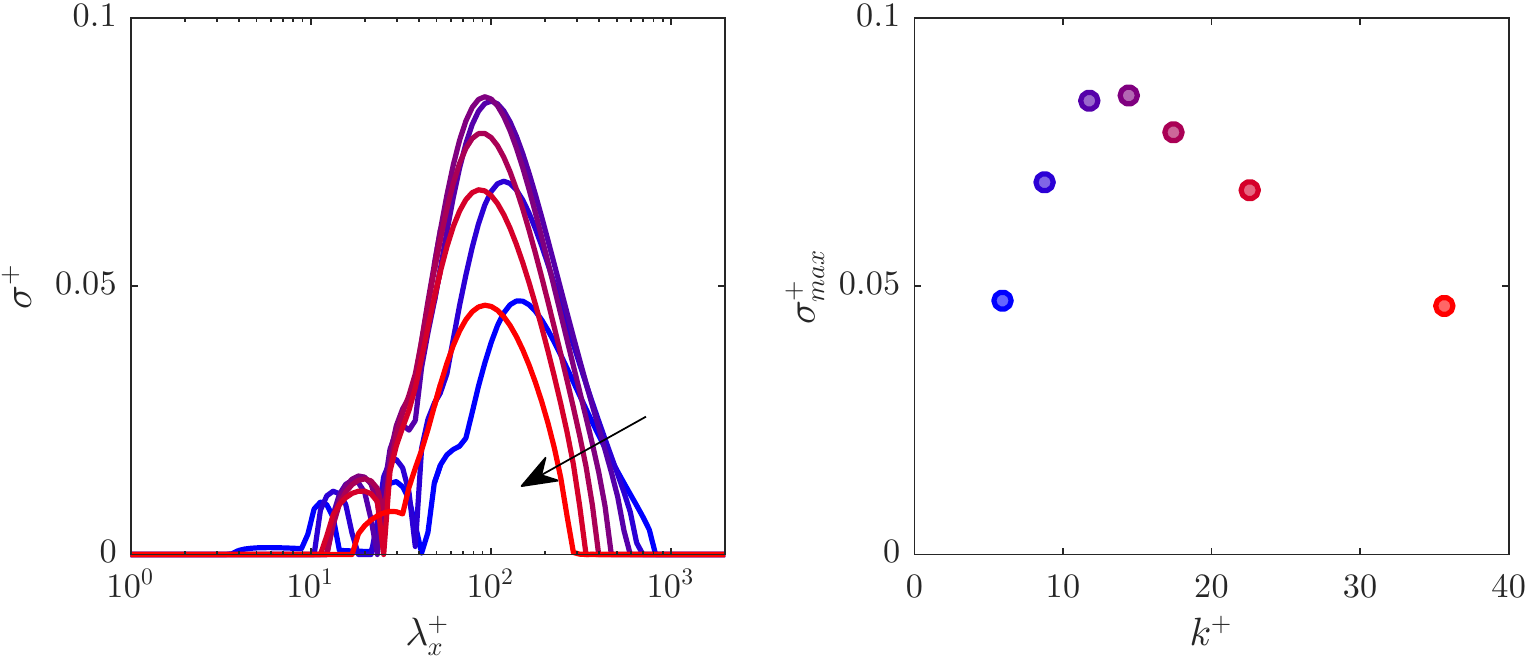}%
\mylab{-0.905\textwidth}{5.25cm}{\aaa)}%
\mylab{-0.390\textwidth}{5.25cm}{\bbb)}%
\caption{(\aaa) Growth rate $\sigma^+ = \text{Im}\left(\omega^+\right)$ of the most amplified mode as a function of the longitudinal wavelength $\lambda_x^+$. (\bbb) Maximum growth rate $\sigma^+$ as a function of the roughness height. Blue to red, results for the mean velocity profiles for cases C06 to C36.}
\label{fig:ampl_vs_lx}
\end{figure}
We focus on Kelvin-Helmholtz-like instabilities, which are essentially spanwise coherent, linear and inviscid.
Previous studies have shown that the instability is essentially a property of the mean profile \citep{Beneddine2016}, although it is modulated by the effect of the complex substrate on the fluctuating flow \citep{Raupach1991,White2007,Zampogna2016b}, as well as by viscous effects \citep{Jim2001,Luminari2016,GG2018}.
Since we aim to assess this phenomenon only qualitatively, we conduct a simplified analysis, two-dimensional in $x$ and $y$, inviscid and linear, where the flow is allow to fluctuate freely around the mean profile, neglecting the presence of the solid obstacles. 
We seek wavelike solutions for the velocity and pressure perturbations, of the form \mbox{$f=\hat{f}\exp[i(\alpha_x x + \alpha_z z - \omega t)]$}, which allows for a modal analysis.
The linearised problem becomes then Rayleigh's equation \citep{Rayleigh1879}
\begin{equation}
  \left((\meanU-c)(\partial_{yy} - k^2) - \partial_{yy}\meanU \right) \hat{v} = 0, 
  \label{eq:rayleigh}
\end{equation}
where $\alpha_x$ and $\alpha_z$ are the wave numbers in the streamwise and spanwise directions respectively, $\hat{v}$ is the corresponding perturbation mode of the wall-normal velocity, with $\hat{v}=0$ at the troughs, $k^2 = \alpha_x^2 + \alpha_z^2$, $\omega$ is the complex frequency, and $c$ is the complex phase velocity defined as $\omega = \alpha_x c$.
Note that, according to Squire's theorem, for any streamwise wavelength $\alpha_x$ the most amplified mode is two-dimensional, $\alpha_z = 0$.
The mean velocity profiles, \meanU, are directly extracted from our DNSs, and include the region below the roughness crests.
Notice that the effect of roughness is exclusively introduced through the mean velocity profile. 
More precise studies include, for instance, a drag force model below the roughness or canopy tips \citep{Py2006,Luminari2016,AK2018}, at the cost of an increased cost and complexity.

The results of the stability analysis, portrayed in figure~\ref{fig:ampl_vs_lx}(\aaa), show an instability for the mean flow predominantly for wavelengths $\lambda_x^+ \approx 100$--$150$.
This result is in rough agreement with the modifications observed in the energy spectra and cospectra in figure~\ref{fig:outer_spectra_u_v_uv_y8}, where energy concentrates at $\lambda_x^+ \sim 150$, and does not significantly change with \kp.
This could be expected, as previous studies have shown that the lengthscales of the instability are set by the curvature of the mean velocity profile, independently of the lenghtscales in the substrate geometry \citep{White2007,GarM2011,GG2018}.
However, our simplified model exhibits a maximum for the instability at $\kp \approx 15$, as depicted in figure~\ref{fig:ampl_vs_lx}(\bbb), while the changes in the energy spectra increase monotonically with \kp.
Nevertheless, the results suggest that the wavelengths in which energy concentrates in the DNSs are the most receptive from the point of view of the stability of the mean flow.

\section{Skin friction: towards a predictive model}\label{sec:stress_breakdown}

\begin{figure}
\centering
\includegraphics[width=.9\textwidth]{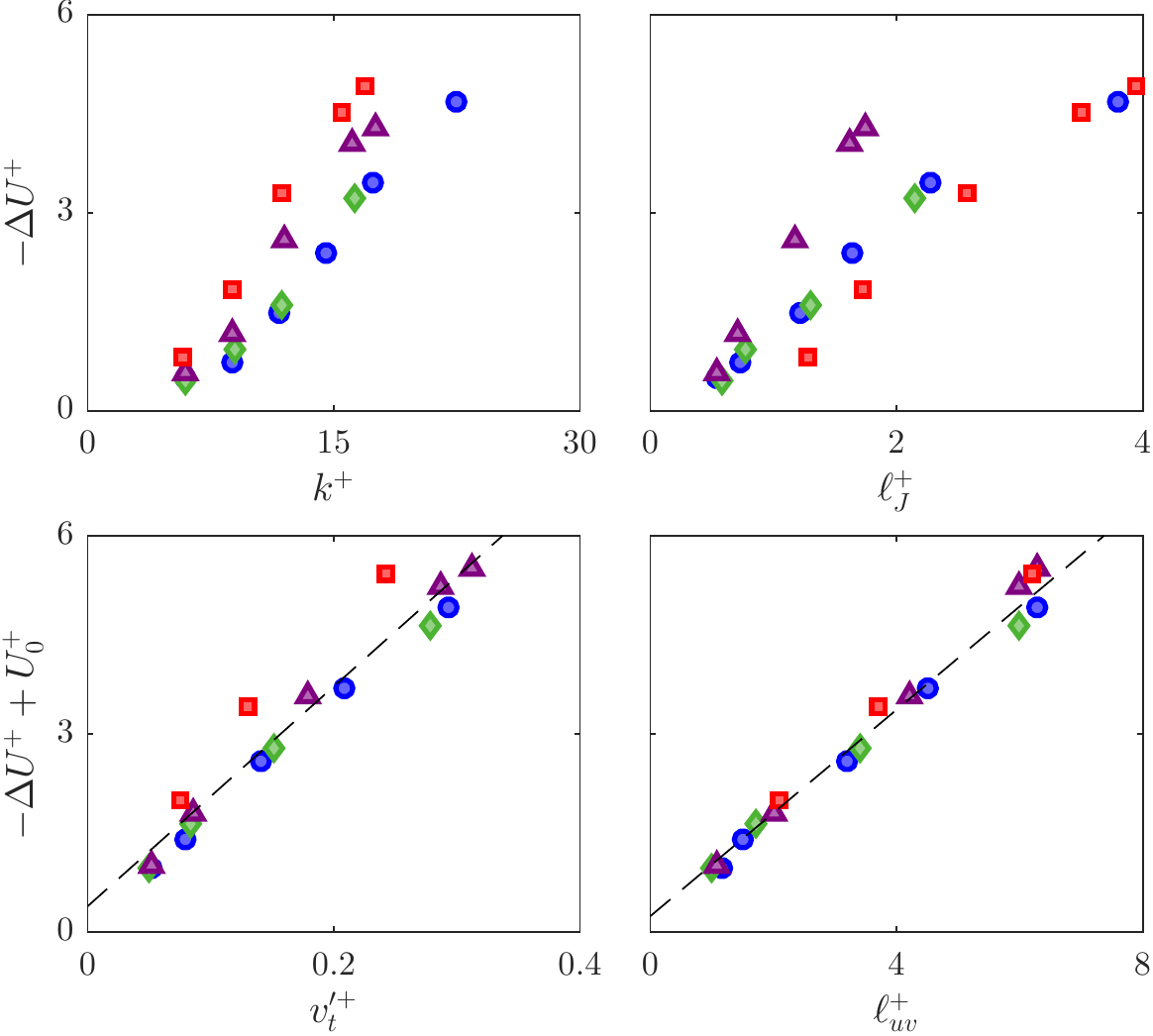}%
\mylab{-0.820\textwidth}{10.40cm}{\aaa)}%
\mylab{-0.380\textwidth}{10.40cm}{\bbb)}%
\mylab{-0.820\textwidth}{4.90cm}{\ccc)}%
\mylab{-0.380\textwidth}{4.90cm}{\ddd)}%
\caption{Roughness function represented versus (\aaa) roughness height, (\bbb) Jackson's displacement height, (\ccc) wall-normal velocity fluctuations at the roughness crests, and (\ddd)~virtual origin of turbulence. \protect\bluecircle, collocated posts; \protect\greendiamond, spanwise-staggered posts; \protect\purpletriangle,~streamwise staggered posts; \protect\redsquare, collocated posts of two heights; \protect\blacklinedash , linear regressions  $16.6\, v_t^{\prime +}$ and  $0.78\, \Duvp$}
\label{fig:DU_vs_kp_and_Duv}
\end{figure}
In the Introduction we discuss the main drawbacks of the equivalent sand roughness, \ksp, which for historical reasons has been widely used to characterise rough surfaces.
However, \ksp\ can neither be predicted a priori, nor describes appropriately the transitionally rough regime.
The latter is explicitly highlighted by \citet{Jim2004}, who shows that a collapse of the fully rough regime does not guarantee such a collapse in the transitionally rough regime, as shown in figure~\ref{fig:transitional_roughness_adapted}.
Therefore, \ksp\ is not a suitable parameter to predict \DUp.
Likewise, \kp\ presents similar problems, as it is roughly proportional to \ksp\ \citep{Schlichting1936}.
Figure~\ref{fig:DU_vs_kp_and_Duv} shows \DUp\ as a function of the roughness element height, \kp, and of Jackson's displacement height, \Djp.
Both capture the trend of increasing \DUp\ for increasing roughness size, but display a strong dependence with the type of roughness surface.
\citet{Orlandi2006a} find a strong linear correlation between $\DUp-\Utipsp$ and the rms fluctuations of the wall-normal velocity at the roughness crests, \vtipsp, as shown in figure~\ref{fig:DU_vs_kp_and_Duv}(\ccc) for our simulations.
Similarly, a linear relationship between $\DUp-\Utipsp$ and \Duvp\ is also observed in figure~\ref{fig:DU_vs_kp_and_Duv}(\ddd).
As introduced earlier, \Duvp\ is the shift of the Reynolds stress below the roughness crests, and can be interpreted as the apparent position of the origin for turbulence.
Both \vtipsp\ and \Duvp\ establish a connection between the roughness function and the effect of the roughness surface on the flow.

The mean momentum equation is used below to explore this relationship between the roughness function, \DUp, and the virtual origin of the Reynolds shear stress, \Duvp.
The goal is to obtain an expression for the roughness function.
The procedure followed here is similar to that in \citet{GarM2011} to study the contributions to \DUp.
In a turbulent channel the mean momentum equation along the streamwise direction is
\begin{align}
  \ReStress + \nu \frac{dU}{dy} &= u^2_{\tau} \frac{\delta'-y_r}{\delta'},
  \label{eq:total_streess_with_dimensions}
\end{align}%
where $y_r$ is the wall-normal coordinate measured from the virtual origin of the mean velocity profile, $y_r = y - \protu$, and the apparent half-height of the channel has previously been defined as $\delta' = \delta + \protu$.
This allows us to define a common origin $y_r = 0$ for the mean velocity profiles under different setup configurations.
Note that $\protu$ is positive and $\delta' > \delta$.
Scaling equation~\eqref{eq:total_streess_with_dimensions} in viscous units gives
\begin{align}
  {\ReStressp} + \frac{dU^{+}}{dy^{+}} &= \frac{\delta^{\prime+}-y_r^{+}}{\delta^{\prime+}}.
  \label{eq:total_streess_with_no_dimensions}
\end{align}
This expression is valid above the roughness crests, $\yp > 0$ or $\yp_r > \protup$.
  Integrating equation~\eqref{eq:total_streess_with_no_dimensions} allows us to obtain $U^+$, and thus an expression for the roughness function, \DUp, as the difference in $U^+$ between rough and smooth-wall cases.
  At a distance $H^+$ sufficiently far from the wall, where outer-layer similarity holds and the mean velocity profile is logarithmic, the roughness function is $\DUp = U^+_r(H^+) + U^+_s(H^+)$.
Let us denote by the subscripts `$r$' and `$s$' the variables in a rough and smooth-wall channel, respectively.
Equation~\eqref{eq:total_streess_with_no_dimensions} can then be integrated between two heights 
  \begin{subequations}
    \begin{align}
      \int^{H^+}_{h_s^+}{\ReStressp[s]} \, dy^+ + U^{+}_s\left(H^+\right) - U^{+}_s\left(h_s^+\right) &= (H^+ - h_s^+) -\frac{1}{2}\frac{{H^+}^2 - {h_s^+}^2}{\delta^{\prime+}_s}
      \label{eq:stress_integrals_smooth},\\ 
      \int^{H^+}_{h_r^+}{\ReStressp[r]} \, dy^+ + U^{+}_r\left(H^+\right) - {U^{+}_r}\left(h_r^+\right) &= (H^+ - h_r^+) -\frac{1}{2}\frac{{H^+}^2 - {h_r^+}^2}{\delta^{\prime+}_r},
      \label{eq:stress_integrals_rough}
    \end{align}
    \label{eq:stress_integrals_smooth_and_rough}%
  \end{subequations}%
where the lower bounds of integration, $h_r^+$ and $h_s^+$, can in principle be different for the rough and the reference smooth case, but for equation~\eqref{eq:total_streess_with_dimensions} to hold, $h_r^+ \geq \protup$ is required.
An expression for the roughness function, $\DUp$, can be obtained by subtracting equation~\eqref{eq:stress_integrals_smooth} from~\eqref{eq:stress_integrals_rough}.
Taking $h^+ = h_s^+ = h_r^+ = \protup$, we then have
\begin{equation}
  \DUp = U^{+}_r\left(H^+\right) -  U^{+}_s\left(H^+\right) = \Tterm_1 + \Tterm_2 + \Tterm_3,
  \label{eq:stress_integrals_DU}
\end{equation}
where
  \begin{subequations}
    \begin{align}
      \Tterm_1 =& -\left( \int^{H^+}_{h^+}{\ReStressp[r]} \, dy^+ 
      - \int^{H^+}_{h^+}{\ReStressp[s]} \, dy^+ \right),
      \label{eq:stress_integrals_A1}\\
      \Tterm_2 =& {U^{+}_r}\left(h^+\right) - {U^{+}_s}\left(h^+\right),
      \label{eq:stress_integrals_A2}\\
      \Tterm_3 =& - \frac{1}{2} \left( \frac{{H^+}^2 - {h^+}^2}{\delta^{\prime+}_r} - \frac{{H^+}^2 - {h^+}^2}{\delta^{\prime+}_s}\right).
      \label{eq:stress_integrals_A3}%
    \end{align}%
    \label{eq:stress_integrals_A}%
  \end{subequations}%

The first term, $\Tterm_1$, encapsulates the increase in Reynolds stress that roughness produces compared to that over a smooth wall.
Over conventional roughness, the Reynolds stress increases near the wall and hence the negative contribution of this term towards \DUp.
Drag-reducing surfaces produce large slip velocities that overcome the effect of $\Tterm_1$, resulting in a net reduction of drag.
Similarly, roughness also displays a mean velocity at the roughness crests, which is captured by the first term in $\Tterm_2$.
However, when the change in the virtual origin of the mean velocity profile is accounted for, by the second term in $\Tterm_2$, the net result is generally negative.
The first contribution is simply the mean velocity at the roughness crests, \Utips.
The other contribution to $\Tterm_2$ is the mean velocity of the smooth-wall channel at the roughness crests equivalent height, that is what the mean velocity would be at the tips if the rough wall had no effect on the mean velocity profile.
The term $\Tterm_2$ then represents the difference between the actual mean velocity at the roughness crests and the ideal velocity that would have been achieved at such height without roughness.
Notice that for small roughness, $\Tterm_2 \approx 0$.
As size increases, the advective terms gain relevance at the roughness crests and the magnitude of $\Tterm_2$ increases.
Finally, $\Tterm_3$ accounts for Reynolds-number discrepancies between the different simulations, being zero if $\delta_r^{\prime+} = \delta_s^{\prime+}$.
In essence, the term $\Tterm_3$ is obtained by integrating the total stress, i.e.\ the linear, right-hand-side of equation~\eqref{eq:total_streess_with_no_dimensions}.
The intersect of the linear total stress at different $\delta^+$ causes relative variations in $\Tterm_3$ of order $\delta^{\prime+}_r/\delta^{\prime+}_s$.

\begin{figure}
\centering
\includegraphics[width=.762\textwidth]{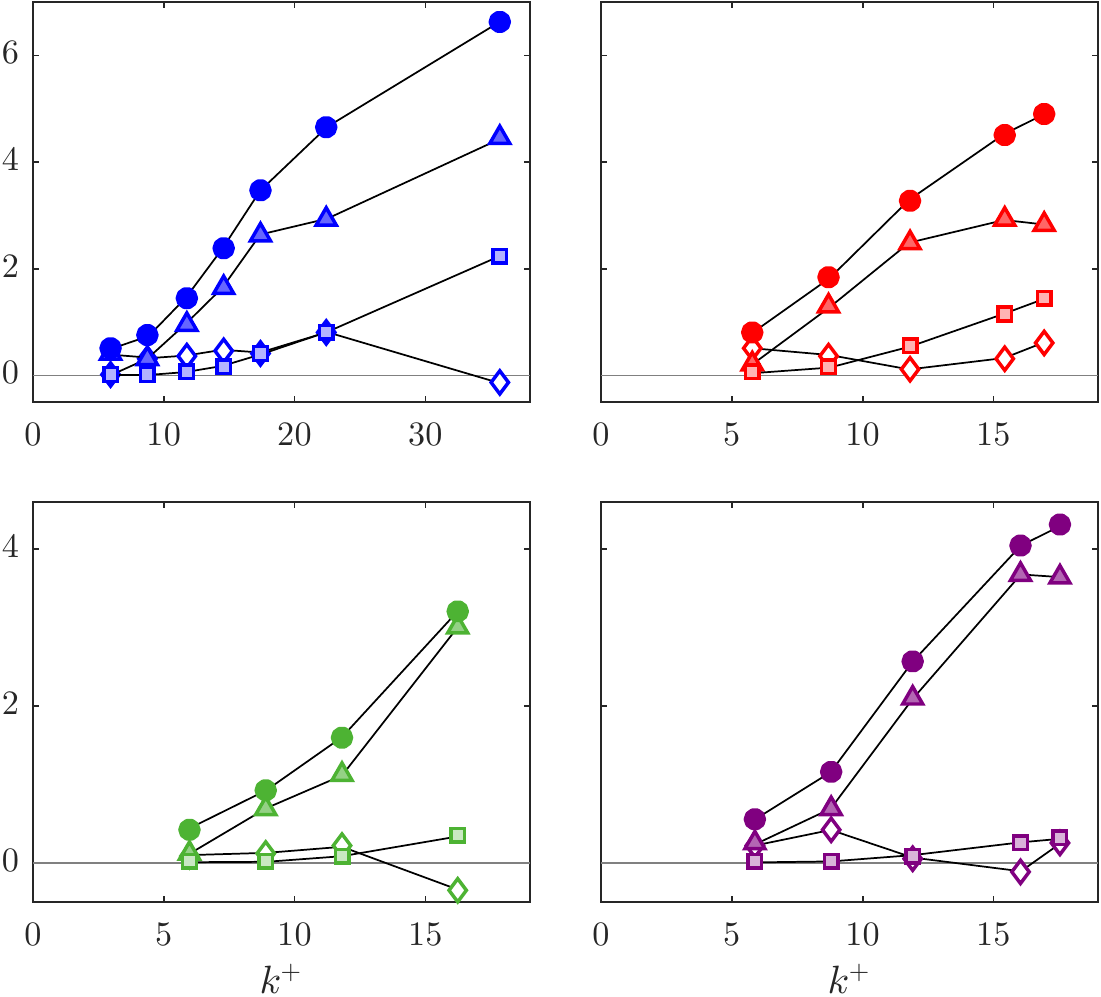}%
\mylab{-.730\textwidth}{.665\textwidth}{\aaa)}%
\mylab{-.335\textwidth}{.665\textwidth}{\bbb)}%
\mylab{-.730\textwidth}{.320\textwidth}{\ccc)}%
\mylab{-.335\textwidth}{.320\textwidth}{\ddd)}%
\caption{Contributions to \DUp\ for (\aaa) collocated posts, (\bbb) collocated posts of two heights, (\ccc) spanwise-staggered posts, and (\ddd) streamwise-staggered posts. \protect\blacktri, $T1$; \protect\blacksquar, $T2$; and \protect\blackdiamond, $T3$; \protect\blackcircle, $\DUp = T1 + T2 +T3$.}
\label{fig:stress_breakdown}
\end{figure}
In our simulations, $\Tterm_3$ contributes significantly to $\DUp$.
The simulations are at slightly different \Retau, which generates setup-dependent contributions to this term.
To isolate these effects, we propose an alternative breakdown, $\DUp = T1 + T2 + T3$, where 
  \begin{subequations}
    \begin{align}
      T1 =& -\left( \frac{\delta_s^{\prime+}}{\delta_r^{\prime+}} \int^{H^+}_{h^+}{\ReStressp[r]} \, dy^+ 
      - \int^{H^+}_{h^+}{\ReStressp[s]} \, dy^+ \right),
      \label{eq:stress_integrals_T1}\\
      T2 =& {U^{+}_r}\left(h^+\right) - {U^{+}_s}\left(h^+\right),
      \label{eq:stress_integrals_T2}\\
      \begin{split}
        T3 =& 
        \left( \frac{\delta_s^{\prime+}}{\delta_r^{\prime+}}-1\right)\int^{H^+}_{h^+}{\ReStressp[r]} \, dy^+
        \underbrace{- \frac{1}{2} \left( \frac{{H^+}^2 - {h^+}^2}{\delta^{\prime+}_r} - \frac{{H^+}^2 - {h^+}^2}{\delta^{\prime+}_s}\right)}_{\Tterm_3}.
      \end{split}%
      \label{eq:stress_integrals_T4}%
    \end{align}%
    \label{eq:stress_integrals}%
  \end{subequations}%
These expressions are obtained by adding and subtracting $\delta^{\prime+}_s/\delta^{\prime+}_r\int \langle uv \rangle_r^+$ to equation~\eqref{eq:stress_integrals_A}, and rearranging.
In this form, the term $|T3| \ll |T1|, |T2|$, and the expression $\DUp = \Tterm_1 + \Tterm_2$ is still recovered for $\delta_r^{\prime+} = \delta_s^{\prime+}$.
By rescaling ${\ReStressp[r]}$ by the ratio of channel heights, $\dsp/\drp$, $T1$ is less dependent to small variations of the frictional Reynolds number, and thus cases with slightly different $\delta^+$ can be more fairly compared.
Results of equations~\eqref{eq:stress_integrals} used on our rough geometries are portrayed in figure~\ref{fig:stress_breakdown}.
The term $T1$ contributes the most towards \DUp.
In our geometries, the term $T2$ is observed to always be negative. 
However, some particular cases with two-dimensional roughness, such as riblets, have also proven to induce a positive, and therefore drag reducing, $T2$ term \citep{GarM2011}.

  For $\delta_r^+ = \delta_s^+$, equations~\eqref{eq:stress_integrals_A} and~\eqref{eq:stress_integrals} simplify to
\begin{subequations}
\begin{align}
  T1 = \mathcal{T}1 =&   \int^{\delta^{\prime+}_s}_{h^+}\left( \minusReStressp[r] - \minusReStressp[s] \right)\, dy^+,\\
  T2 = \mathcal{T}2 =& {U^{+}_r}\left(h^+\right) - {U^{+}_s}\left(h^+\right).%
\end{align}%
\label{eq:stress_integrals_simple}%
\end{subequations}%
Equations~\eqref{eq:stress_integrals_simple} present in a clearer manner the components of \DUp, as it is caused by the change in Reynolds stress as well as the difference between the mean velocity at the roughness crests and that over a reference smooth wall at the same height from the virtual origin.

\begin{figure}
\centering
\includegraphics[width=.97\textwidth]{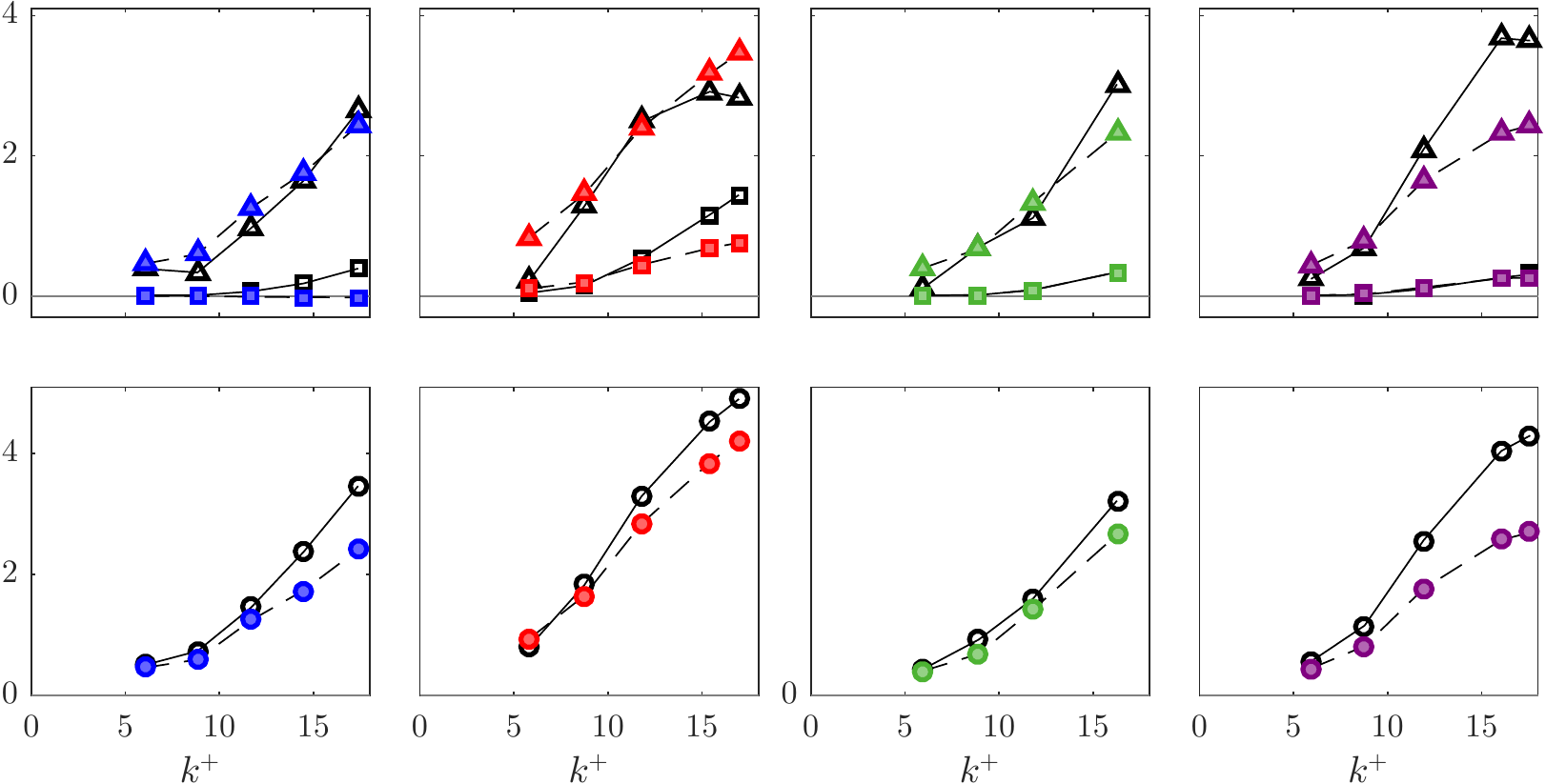}%
\mylab{-.940\textwidth}{.465\textwidth}{\aaa.1)}%
\mylab{-.695\textwidth}{.465\textwidth}{\bbb.1)}%
\mylab{-.452\textwidth}{.465\textwidth}{\ccc.1)}%
\mylab{-.205\textwidth}{.465\textwidth}{\ddd.1)}%
\mylab{-.940\textwidth}{.225\textwidth}{\aaa.2)}%
\mylab{-.695\textwidth}{.225\textwidth}{\bbb.2)}%
\mylab{-.452\textwidth}{.225\textwidth}{\ccc.2)}%
\mylab{-.205\textwidth}{.225\textwidth}{\ddd.2)}%
\caption{Results from DNSs and estimates based on equations~\eqref{eq:stress_integrals_simple}. \protect\blacktri,~$\Tterm_1$; \protect\blacksquar,~$\Tterm_2$; \protect\blackcircle,~\DUp. (\aaa) Collocated posts; (\bbb) collocated posts of two heights; (\ccc) spanwise-staggered posts; and (\ddd) streamwise-staggered posts. Solid symbols, estimates; empty symbols, DNSs.}
\label{fig:stress_breakdown_2}
\end{figure}
Now we explore the potential of equations~\eqref{eq:stress_integrals_simple} for predicting \DUp.
Based on the discussion in \S\ref{sec:rough_comp} and \S\ref{sec:the_turbulent_component}, we suggest a model for the contributions $\Tterm_1$ and $\Tterm_2$ to estimate \DUp.
Let us assume that the statistics for a turbulent flow over a smooth wall at the desired \Retau\ are available.
Therefore, only the terms from roughness, $U_r^+(h^+)$ and $\minusReStressp[r](\yp)$, need to be modelled.
For small roughness size, the velocity at the roughness crests, $U_r^+(h^+)$, can be estimated from the laminar model for the coherent flow presented in \S\ref{sec:rough_comp}.
In \S\ref{sec:the_turbulent_component} it is shown how for small \kp\ the main effect of roughness on the Reynolds stress, \minusReStressp[r], is as a shift \Duvp\ towards the wall, but otherwise the Reynolds stress closely resembles that of smooth-wall turbulence.
As a result the effective displacement of the Reynolds stress is $\Duvp - \protup$, since the origin is at a depth \protup\ below the roughness tips.
We can express this relation as if the Reynolds stress of a rough wall was that of smooth-wall turbulence shifted to the corresponding virtual origin, i.e.\ $\minusReStressp[r](\yp) \approx \minusReStressp[s](\yp_\star)$, where $\yp_\star$ is an auxiliary wall-normal coordinate that displaces \minusReStressp[s].
This auxiliary coordinate $\yp_\star$ is defined such that near the wall, at $\yp_r = h^+$, $\minusReStressp[r](\yp_r = h^+) \approx \minusReStressp[s](\yp_r = h^+ + \Duvp - \protup)$.
Notice that a mere shift of \minusReStressp[s] leads to a new $\delta^{\prime +}$. 
Instead, to keep $\delta^{\prime +}$ constant, $\yp_\star$ is linearly transformed, with $\minusReStressp[r](\yp_r = \delta^{\prime+}) \approx \minusReStressp[s](\yp_r = \delta^{\prime+})$.
The change in Reynolds stress, accounted for in the term $\Tterm_1$, can therefore be expressed as
\begin{equation}
  \Tterm_1 \approx \int^{\delta^+}_{h^+}\left( \minusReStressp[s](\yp_\star) - \minusReStressp[s](\yp) \right)\, dy^+,\\
  \label{eq:stress_integrals_T1_simplified}
\end{equation}
where 
\begin{equation}
  \yp_\star = \left(\frac{\delta^{\prime+}-\Duvp}{\delta^{\prime+}-\protup         }\right)\left(\yp_r - \protup\right) + \Duvp.
\end{equation}%
Results of this model are portrayed in figure~\ref{fig:stress_breakdown_2}.
In addition, the values of \hrp\ and \Utips, used to estimate $\Tterm_2$, are estimated from the laminar model for the roughness-coherent contribution, presented in \S\ref{sec:rough_comp}.
The model appears to estimate the initial trend of \DUp\ with relatively good agreement up to values of $-\DUp\lesssim 2$, i.e. capturing the region of initial increase of drag with $\DCf/C_f \lesssim 25\%$.
For larger roughness, the results begin to deviate more significantly.
These results show the potential of the model described by equations~\eqref{eq:stress_integrals_simple} and~\eqref{eq:stress_integrals_T1_simplified} to estimate \DUp\ as it departs from the hydraulically smooth regime.
Further work needs to be undertaken in order to extend the model to the entire transitionally rough regime, as well as to obtain an estimate for \Duvp.
Additionally, its applicability to random roughness must be analysed, especially for the purpose of industrial applications.

\section{Conclusions}\label{sec:conclusions}

In the present work we have investigated the interaction between roughness and near-wall turbulence.
We have focused on the transitionally rough regime, to capture the effects that trigger the departure from the hydraulically smooth regime.
We propose a triple decomposition  where the roughness-coherent contribution is modulated in amplitude by the overlying background-turbulent flow.
Using this decomposition, a background turbulence component, essentially free of any footprint from the roughness texture, can be extracted.

The roughness-coherent component, resulting from the ensemble average of the flow field, is systematically studied as roughness size increases.
This component presents an exponential decay with $y/\k$.
The rate of decay depends on the velocity component, with the wall-normal velocity experiencing the slowest decay.
All components seem to essentially vanish for $y \lesssim \k$.
However, the region where background turbulence is affected by the roughness surface, the roughness-sublayer, extends to a height of \mbox{$y/\k \approx 2$--$3$}.
A laminar model is proposed to estimate the roughness-coherent component.
The model results agree well with DNS results for roughness sizes that produce offsets of $\DUp \lesssim 2$.
It also provides estimates for the mean velocity at the roughness crests, \Utipsp, and the virtual origin of the mean velocity profile, \protup.

Using the proposed triple decomposition, the roughness-coherent contribution is extracted and the changes produced in the background turbulence can be isolated and analysed.
For a large extent of the transitionally rough regime, $\DUp \lesssim 4$, the main effect of roughness is a displacement of the rms fluctuations and the Reynolds stress towards the wall.
This is interpreted as turbulence perceiving the wall at a certain virtual origin below the roughness crests.
We observe a good agreement between \protup, \Dup\ and \Djp, the origins perceived by the mean flow, the streamwise velocity fluctuations and the displacement height, respectively.
This supports Jackson's methodology for defining the displacement height for the law of the wall.
However, the near-wall turbulence appears to experience a different virtual origin, essentially that of the Reynolds stress, \Duvp.
This origin presents a strong correlation with \DUp, similar to that observed for the wall-normal velocity fluctuations by \citet{Orlandi2006a}.

As the roughness size increases and the effect on the background turbulence ceases to be a mere shift, the spectral energy density increases at short streamwise wavelengths, especially for the wall-normal velocity.
This increase tends to concentrate at $\lambda_x^+ \approx 150$.
Results from a simplified linear stability model suggest that this concentration is related to an increased receptivity for $\lambda_x^+ \approx 100$--$150$, produced by the inflexionality of the mean velocity profile.

The mean momentum equation has been integrated to identify different contributions to the roughness function, \DUp, and in particular to analyse the role of the virtual origin \Duvp.
The main contribution is the change in Reynolds stress. 
For small roughness, this is essentially due to the shift \Duvp, as the Reynolds stress remains otherwise smooth-wall-like.
The second contribution to \DUp\ is proportional to the defect of the mean velocity at the height of the roughness crests, compared to a smooth-wall flow. 
This defect is generally positive, in which case it increases drag.\\

N.A-E.\ and C.T.F.\ were supported by the Engineering and Physical Sciences Research Council under a Doctoral Training Account, grant EP/M506485/1. 
This work was partly supported by the European Research Council through the II Multiflow Summer Workshop. 
We also express our gratitude to A.~Sharma for helpful conversations about shear-flow instabilities and their analysis.
\appendix
\section{}\label{appendixA}
Approximate decomposition of the rms fluctuations using equations~\ref{eq:flow_decomposition}
\begin{subequations}
  \begin{align}
  \langle u^{\prime 2} \rangle 
  =& \left< \uT^{2} \right> 
  + \left< \uRu^2 \right> 
  + \left< \uRu^{2} \right> \left< \frac{\uT^{2}}{\meanU^2} \right> 
  + \left< \uRv^{2} \right> \left< \frac{\vT^{2}}{\tilV^2} \right>
  + \left< \uRw^{2} \right> \left< \frac{\wT^{2}}{\tilW^2} \right> \nonumber \\ 
  +& 2\left(\left< \uRu \uRv \right> \left< \frac{\uT \vT}{\meanU \tilV} \right> 
  + \left< \uRu \uRw \right> \left< \frac{\uT \wT}{\meanU \tilW} \right>
  + \left< \uRv \uRw \right> \left< \frac{\vT \wT}{\tilV \tilW} \right> \right)
  \end{align}
  \begin{align}
  \langle v^{\prime 2} \rangle 
  =& \left< \vT^{2} \right> 
  + \left< \vRu^2 \right> 
  + \left< \vRu^{2} \right> \left< \frac{\uT^{2}}{\meanU^2} \right> 
  + \left< \vRv^{2} \right> \left< \frac{\vT^{2}}{\tilV^2} \right>
  + \left< \vRw^{2} \right> \left< \frac{\wT^{2}}{\tilW^2} \right> \nonumber \\ 
  +& 2\left(\left< \vRu \vRv \right> \left< \frac{\uT \vT}{\meanU \tilV} \right> 
  + \left< \vRu \vRw \right> \left< \frac{\uT \wT}{\meanU \tilW} \right>
  + \left< \vRv \vRw \right> \left< \frac{\vT \wT}{\tilV \tilW} \right> \right)
  \end{align}
  \begin{align}
  \langle w^{\prime 2} \rangle 
  =& \left< \wT^{2} \right> 
  + \left< \wRu^2 \right> 
  + \left< \wRu^{2} \right> \left< \frac{\uT^{2}}{\meanU^2} \right> 
  + \left< \wRv^{2} \right> \left< \frac{\vT^{2}}{\tilV^2} \right>
  + \left< \wRw^{2} \right> \left< \frac{\wT^{2}}{\tilW^2} \right> \nonumber \\ 
  +& 2\left(\left< \wRu \wRv \right> \left< \frac{\uT \vT}{\meanU \tilV} \right> 
  + \left< \wRu \wRw \right> \left< \frac{\uT \wT}{\meanU \tilW} \right>
  + \left< \wRv \wRw \right> \left< \frac{\vT \wT}{\tilV \tilW} \right>\right)
  \end{align}
  \begin{align}
  \langle u^{\prime} v^{\prime} \rangle 
  =& \left< \uT \vT \right> 
  + \left< \uRu \vRu \right> \nonumber \\
  +& \left< \uRu \vRu \right> \left< \frac{\uT^{2}}{\meanU^2} \right> 
  + \left< \uRv \vRv \right> \left< \frac{\vT^{2}}{\tilV^2} \right>
  + \left< \uRw \vRw \right> \left< \frac{\wT^{2}}{\tilW^2} \right> \nonumber \\ 
  +& \left( \left< \uRu \vRv \right> + \left< \uRv \vRu \right> \right) \left< \frac{\uT \vT}{\meanU \tilV} \right> 
  +  \left( \left< \uRu \vRw \right> + \left< \uRw \vRu \right> \right) \left< \frac{\uT \wT}{\meanU \tilW} \right> \nonumber \\
  +& \left( \left< \uRv \vRw \right> + \left< \uRw \vRv \right> \right) \left< \frac{\vT \wT}{\tilV \tilW} \right> 
  \end{align}%
\end{subequations}

\bibliographystyle{jfm}

\bibliography{refs}

\end{document}